\documentclass{aa}
\usepackage[varg]{txfonts}
\usepackage{natbib}
\usepackage{graphicx}
\usepackage{dcolumn,url,longtable}
\usepackage{amssymb}
\usepackage{color}
\usepackage{tikz}
\usetikzlibrary{shapes,arrows}
\newcolumntype{d}{D{.}{.}{-1}}

\begin{document} 

  \title{Reconstruction of asteroid spin states from Gaia DR3 photometry}
  

  \author{J. \v{D}urech         \inst{1}        \and
          J. Hanu\v{s}          \inst{1}        
         }

  \institute{Charles University, Faculty of Mathematics and Physics, Institute of Astronomy, V Hole\v{s}ovi\v{c}k\'ach 2, 180\,00 Prague, Czech Republic\\
             \email{durech@sirrah.troja.mff.cuni.cz}
             }

  \date{Received ?; accepted ?}

  \abstract
  {}
  {Gaia Data Release 3 contains accurate photometric observations of more than 150,000 asteroids covering a time interval of 34 months. With a total of about 3,000,000 measurements, a typical number of observations per asteroid ranges from a few to several tens. We aimed to reconstruct the spin states and shapes of asteroids from this dataset.}
  {We computed the viewing and illumination geometry for each individual observation and used the light curve inversion method to find the best-fit asteroid model, which was parameterized  by the sidereal rotation period, the spin axis direction, and a low-resolution convex shape. To find the best-fit model, we ran the inversion for tens of thousands of trial periods on interval 2--10,000\,h, with tens of initial pole directions. To find the correct rotation period, we also used a triaxial ellipsoid model for the shape approximation.}
  {In most cases the number of data points was insufficient to uniquely determine the rotation period. However, for about 8600 asteroids we were able to determine the spin state uniquely together with a low-resolution convex shape model. This large sample of new asteroid models enables us to study the spin distribution in the asteroid population. The distribution of spins confirms previous findings that (i) small asteroids have poles clustered toward ecliptic poles, likely because of the  YORP-induced spin evolution, (ii) asteroid migration due to the Yarkovsky effect depends on the spin orientation, and (iii) members of asteroid families have the sense of rotation correlated with their proper semimajor axis:  over the age of the family, orbits of prograde rotators evolved, due to the  Yarkovsky effect, to larger semimajor axes, while those of retrograde rotators drifted in the opposite direction.}
  {}

  \keywords{Minor planets, asteroids: general, Methods: data analysis, Techniques: photometric}

  \maketitle

  \section{Introduction}

The first photometric measurements of asteroids from the ESA Gaia mission \citep{Gaia:16} were released in April 2018 as part of the Gaia Data Release 2 \citep[DR2,][]{Gaia:18}. That dataset contained astrometry and photometry for $\sim$14,000 asteroids \citep{Spo.ea:18} covering a time interval of about 22 months. The recent Gaia Data Release 3 from June 2022 \citep[DR3,][]{Tan.ea:22, Bab.ea:22} provided a significantly larger number of asteroid measurements: more than 3,000,000 photometric data points for about 150,000 asteroids covering a time interval of about 34 months. Typically there are fewer than 20 individual measurements per asteroid in DR3. The largest number of measurements is 96. 

Although the Gaia DR2 photometric dataset was rather limited, we successfully derived physical models for almost 200 asteroids, consisting of mostly new solutions, and we published the results in \citet{Dur.Han:18}. We used the standard convex inversion method \citep{Kaasalainen2001b, Kaa.Tor:01} based on the inversion of photometric measurements. This physical model included the sidereal rotation period, the orientation of the spin axis, and a convex 3D shape model. 

Thanks to a significantly larger amount of data and more extended temporal coverage of the DR3 data compared to DR2, we were expecting a dramatic increase in the number of successful shape model determinations leading to an unprecedented insight into the distribution of physical properties of asteroids. This paper describes our application of the same procedure as in \citet{Dur.Han:18} to process and analyze the DR3 data. 

In Sect.~\ref{sec:method} we describe the DR3 data downloading and processing and the inversion technique we applied, including the verification tests. In Sect.~\ref{sec:results} we present our analysis of the spin properties of asteroids. We conclude our work in Sect.~\ref{sec:conclusions}.

  \section{Inversion of Gaia asteroid photometry}
  \label{sec:method}
  
  We downloaded the data from the Gaia archive.\footnote{\url{https://gea.esac.esa.int/archive/}} We selected only the relevant parameters: the Gaia-centric JD in TCB ({\tt epoch}), the calibrated G-band magnitude ({\tt g\_mag}), the G flux ({\tt g\_flux}), and the error in the G flux ({\tt g\_flux\_error)}. Because the JD epoch is given for each CCD position but the magnitude is the same over the transit, we averaged the epoch values over the transit.\footnote{\tt SELECT number\_mp, avg(epoch), g\_mag, g\_flux\_error / g\_flux AS error FROM gaiadr3.sso\_observation WHERE g\_mag IS NOT NULL GROUP BY number\_mp, g\_mag, error} We converted JD from the original Gaia-centric TCB to TDB according to the formula given in \cite{Tan.ea:22}. We also computed the relative flux error as {\tt g\_flux\_error / g\_flux}. In this way we obtained 3,069,170 photometric data points for 156,789 asteroids (\citealt{Tan.ea:22} report 156,801 in their Table~1 because some asteroids have only NULL magnitudes). Then we computed the heliocentric and Gaia-centric ecliptic coordinates for each observation using the API interface of the JPL Horizons ephemeris service.\footnote{\url{https://ssd.jpl.nasa.gov/horizons/}} We computed the light-time correction, corrected the brightness to 1\,au distance from the Sun and Gaia, and converted magnitudes to relative flux. 
  
\begin{figure}
    \includegraphics[width=\columnwidth]{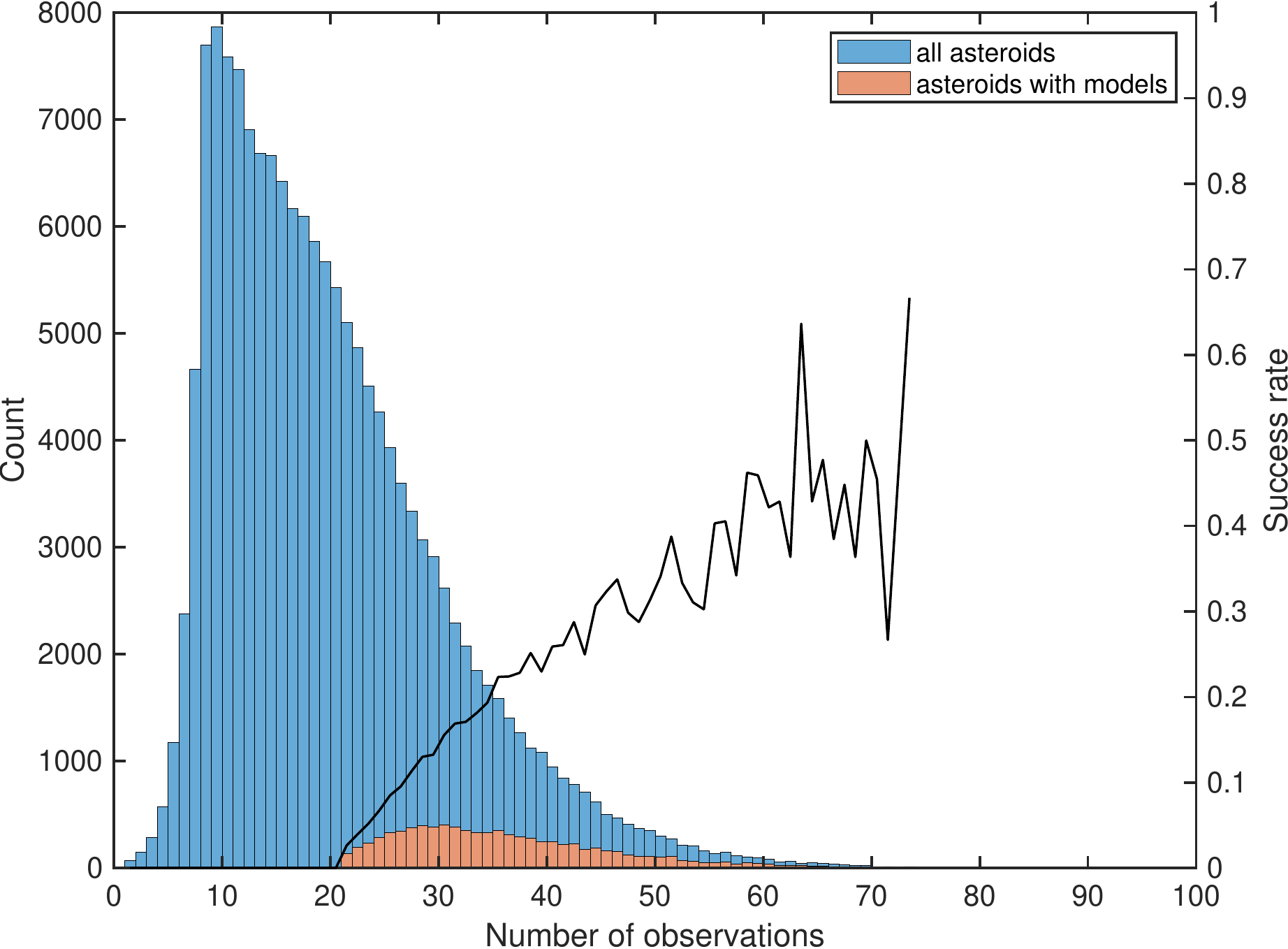}
    \caption{Histogram showing number distributions of  observations per asteroid in the whole DR3 dataset (blue) and for asteroids for which we derived a spin and shape model (orange). The black curve shows the success rate of deriving a model for each data bin with at least ten counts, which is the ratio of the orange to blue bins.}
    \label{fig:histogram}
  \end{figure}

  We limited our sample to 60,945 asteroids that had 21 or more observations. For asteroids with fewer data points, the number of model parameters would be higher than the number of observations, which would often lead to unrealistic fits with zero residuals. The distribution of the number of observations per asteroid is shown in Fig.~\ref{fig:histogram}; the mean number of observations is 20, and the median is 17. There are about 2000 asteroids with more than 50 data points, 473 with more than 60, and only 89 with more than 70.
  
  We performed a period search with a convex shape model parameterized by spherical harmonics on the order and degree of three. The shape was discretized with 288 surface elements of areas $\sigma_i$ with normals $\vec{n_i}$ isotropically distributed in spherical coordinates $\vartheta$ and $\varphi$. Instead of directly optimizing  $\sigma_i$, we used the expansion into spherical harmonics series with associated Legendre polynomials $P_l^m$
  \begin{equation}
  \sigma_i(\vartheta_i, \varphi_i) = \sum_{l = 0}^3 \sum_{m=-l}^l \left(a_{lm} \cos m\varphi_i + b_{lm} \sin m \varphi_i\right) P_l^m(\cos\vartheta_i)\,,
  \end{equation}
  where $a_{lm}$ and $b_{lm}$ were subject to optimization. The parameter $a_0$  corresponds to the size of the shape model. Because we do not know the albedo, the size cannot be determined and we set $a_0$ to some arbitrary fixed value and only the remaining 15 $a_{lm}$, $b_{lm}$ parameters are optimized. Periodograms were computed with the errors of individual photometric measurements taken into account. To avoid  errors that are too small and that  are below the resolution of the convex-shape model, we set up the minimum relative error to 0.01.

\begin{figure}
    \includegraphics[width=\columnwidth]{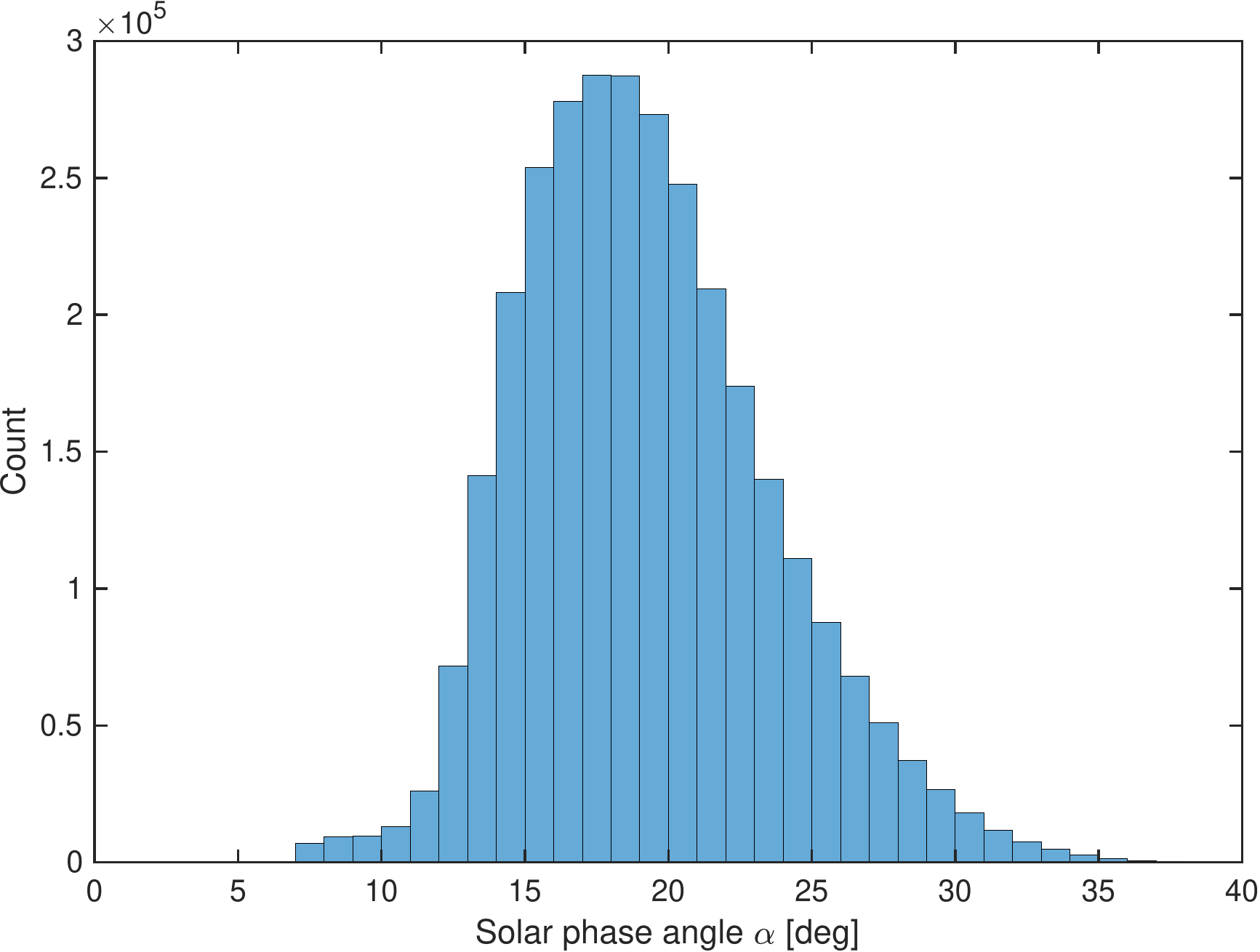}
    \caption{Histogram showing      solar phase angle $\alpha$ distributions for all asteroid observations.}
    \label{fig:histogram_phase}
  \end{figure}

  An asteroid's rotation state was described by its spin axis direction in ecliptic coordinates $(\lambda, \beta)$ and the sidereal rotation period $P$.
  The last parameter of our model was the slope $k$ of the phase curve. For the light scattering model we used a combination of Lambert $S_\text{L} = \mu\mu_0$ and Lommel-Seeliger $S_\text{LS} = \mu\mu_0 / (\mu + \mu_0)$ multiplied by a phase function $f(\alpha)$,  $\mu_0$ and $\mu$ being cosines of the angles of incidence and reflection, respectively. So the scattering model was
  \begin{equation}
  S(\mu, \mu_0, \alpha) = f(\alpha) \left[ S_\text{LS}(\mu, \mu_0) + c\, S_\text{L}(\mu, \mu_0)\right]\,,
  \end{equation}
  where $c$ was fixed at $0.1$ and the phase function had the form \citep{Kaa.ea:02c}
  \begin{equation}\label{eq:phase}
      f(\alpha) = a_0\exp\left(-\frac{\alpha}{d}\right) + k\alpha + 1\,.
  \end{equation}
  The distribution of solar phase angles for all DR3 asteroid observations is shown in Fig.~\ref{fig:histogram_phase}. Most of the observations were carried out at phase angles between 10 and 30\,deg where a linear function can approximate the phase curve. Therefore, we optimized only the linear parameter $k$ in eq.~(\ref{eq:phase}), while the other two parameters describing the increase in brightness near opposition were fixed at typical values $a_0 = 0.5$, $d = 0.1$. The values of $k$ for the sample of reconstructed models (Sect.~\ref{sec:final_models}) were distributed between $-1.5$ and $0$ with the mean value $-0.88$ and standard deviation $0.20$. The shape of the distribution was close to the Gaussian distribution.
    
  The periodograms were computed on an interval of 2--10,000\,h. The step size in the period was set to $0.8\,\Delta P$, where $\Delta P = 0.5\,P^2/T$ is a typical distance between local minima in the period \citep{Kaa:04}, which means that  the time step increases quadratically with increasing period.  In frequencies this corresponds to a uniform sampling with the step $0.4 / T$. For the longest Gaia DR3 dataset with $T \approx 34$ months, $\Delta P \approx 6 \times 10^{-5}$\,h for $P = 2$\,h and $\Delta P \approx 1600$\,h for the longest period of 10,000\,h. The total number of period steps on the whole interval was tens of thousands. A globally unique period solution was defined as having the lowest $\chi^2_\text{min}$ with all other periods giving $\chi^2$ higher than $\chi^2_\text{tr} = (1 + \sqrt{2/\nu})\,\chi^2_\text{min}$, where $\nu$ is the number of degrees of freedom, and the root mean square (RMS) residuals of all local minima had to be higher than 0.01. We obtained a sample of 14,192 asteroids with a unique period defined this way. Then we performed a pole search with the same $\chi^2$ limit and selected only asteroids with one or two pole solutions, which  reduced the sample to 11,854 asteroids. Starting from these poles, we created final models and reconstructed their 3D shape by the Minkowski procedure \citep{Lam.Kaa:01}. We selected only asteroids for which this conversion was successful and for which the ratio of their moment of inertia along the principal axis to that along the actual rotation axis was less than 1.1 \citep{Dur.ea:16}, otherwise the shape would be unrealistically elongated along its rotation axis. This resulted in shape models for 8820 asteroids. We further rejected all asteroids with two pole solutions that had differences in pole latitudes larger than 50\,deg and with differences in longitudes smaller than 120\,deg \citep{Dur.ea:16}. These limits were set arbitrarily to filter out suspicious solutions with pairs of poles that were too far from the expectation that the ambiguity in the spin axis direction leads to two poles with the same latitudes and longitudes difference of 180\,deg \citep{Kaa.Lam:06}. The number of asteroids then reduced to 8230.
  
  As in \citet{Dur.Han:18}, we also used a model of a geometrically scattering triaxial ellipsoid to construct the periodograms The step in the period was set to $0.5\,\Delta P$ in this case. The advantage of this simpler shape model approximation is that the number of model parameters is smaller than in the case of convex models (only two parameters describing the shape:  semiaxes $a$ and $b$ with the shortest axis normalized to unity), the shape model always rotates in a physically correct way (the ellipsoid rotates along its shortest axis), and false half-period solutions are absent. When assuming that the scattering is geometric, the disk-integrated brightness of a triaxial ellipsoid under a general viewing and illumination geometry is proportional to the projected area of the visible and illuminated surface, which can be computed analytically \citep{Ost.Con:84}. It makes this approach about a hundred times faster than when using convex shapes. However, ellipsoidal shape models are insufficient when an asteroid's light curve significantly differs from a simple double sinusoidal curve. The analysis of ellipsoid-based periodograms resulted in 16,010 unique periods. After finding the unique period, we modeled the data with the standard convex shape approximation and applied the same selection criteria described above. We derived solutions with one or two poles for 9910 asteroids, 7873 of which had physically plausible inertia tensors, and 7122 passed the check for the longitude and latitude difference between two pole directions. 

  \subsection{Selecting stable period solutions}
  \label{sec:stable_solutions}
  
  To detect and reject solutions that are not stable with respect to small perturbations of input photometric data, we performed a similar test to that performed in \citet{Dur.Han:18}. We randomly divided each dataset into ten parts, each part containing about one-tenth of the data points. We then removed one part, so about 10\% of data, and repeated the period search and determined the period with the lowest $\chi^2$. We repeated this ten times, removing  10\% of the points in each run and obtaining ten best-fit periods. We then compared these ``jackknife'' periods with the original best-fit period and selected only asteroids for which all ten jackknife periods were the same (within 0.1\%) as the original period or for which there was at most one disagreement between the periods (i.e. there were nine jackknife periods that were  the same as the original). After this selection, we obtained a set of 6269 models based on a convex-shape period search and 5784 models based on a period search with ellipsoids. 
   
  There was some overlap between these two sets. For 3431 asteroids, we had a period solution from convex shape models and ellipsoids. Among them we found 20 asteroids for which the periods were different;  we excluded them from   further analysis, so the final set contained 8602 reliable models.
  
  \subsection{Comparison with Lightcurve Database}
  
  For 3690 asteroids in  our sample, there was an independent period estimate in the Lightcurve Database \citep[LCDB, version from 14 Dec 2021,][]{War.ea:09} that could be compared with our results. To consider only reliable LCDB periods, we selected asteroids with U = 3 uncertainty codes (848 cases). We identified 20 cases where the periods differed by more than 10\%. They are listed in Table~\ref{tab:inconsistent_LCDB_periods}, where we also comment on each case.
  
  Based on our inspection of original publications from which LCDB periods were compiled, we concluded that only five periods were incorrectly determined from Gaia DR3 data. With 848 periods in total, this represents a false solution rate of just $\sim$0.6\%. We removed the five incorrect solutions from our dataset.

      \begin{table*}[t]
      \tiny
    \caption{Asteroids whose rotation period derived from DR3 was different from that in the LCDB.}    \label{tab:inconsistent_LCDB_periods}

    \begin{tabular}{r@{\ }l d d c c p{9cm}}
    \hline
    \multicolumn{2}{c}{Asteroid}  & \multicolumn{1}{c}{$P_\text{Gaia}$}   & \multicolumn{1}{c}{$P_\text{LCDB}$}   & $N$   & Method & Comment on $P_\text{Gaia}$  \\
                            &     & \multicolumn{1}{c}{[h]}               & \multicolumn{1}{c}{[h]}               &       &        &   \\
    \hline    
    197 & Arete                 & 3.15950         & 6.6084    & 33    & E & incorrect  \\ 
    219 & Thusnelda             & 4.44300         & 59.74     & 24    & E & incorrect \\ 
    712 & Boliviana             & 23.463         & 11.743    & 54    & E &    agrees with \cite{Pal.ea:20} \\
    954 & Li                   & 14.4099         & 7.207     & 48    & E &period around 14\,h reported also at Behrend's web page (1) \\ 
    1444 & Pannonia              & 6.9540           & 10.756    & 48    & CE & the same as in \cite{Dur.ea:19}; $P_\text{LCDB}$ is from \cite{Bem.ea:02}, their folded light curve has three maxima per rotation, which is unlikely \\
    1786 & Raahe                 & 30.173          & 18.72     & 33    & CE &  the same as in \cite{Dur.ea:19} and \cite{Dur.Han:18}, also agrees with Behrend's web page (1)\\
    2277 & Moreau                & 17.7253          & 5.397     & 59    & C & likely incorrect; $P_\text{LCDB}$ reported by Pravec is reliable\\
    2760 & Kacha                 & 26.524          & 13.        & 43    & C & \cite{War.Ste:21b} give the same period \\
    3422 & Reid                  & 3.21826           & 2.91      & 64    & CE & agrees with \cite{Dur.ea:19} and \cite{Pal.ea:20} \\
    3507 & Vilas                 & 4.75499           & 3.959     & 25    & C & agrees with \cite{Dur.ea:20} and \cite{Era.ea:20} \\        
    3728 & IRAS                  & 7.0887           & 8.323     & 65    & C & agrees with \cite{Pal.ea:20} \\
    3974 & Verveer               & 13.2437          & 8.51      & 22    & E & agrees with \cite{Dur.ea:20} \\
    4266 & Waltari               & 7.4622          & 11.2      & 40    & C & $P_\text{LCDB}$ reported by \cite{Lec.ea:04} might be incorrect -- the folded light curve is not smooth\\ 
    5436 & Eumelos               & 21.2689         & 38.41     & 32    & CE & agrees with \cite{Sza.ea:17}, \cite{Rya.ea:17}, and \cite{Dur.ea:19}\\ 
    11087 & Yamasakimakoto       & 6.27957           & 4.537     & 27    & C & agrees with \cite{Pal.ea:20} \\
    19562 & 1999 JM81            & 9.0249           & 33.53     & 27    & E & agrees with \cite{Fer:21}; incorrect value in the LCDB by mistake \\
    26858 & Misterrogers         & 12.1225          & 8.065     & 39    & C & period commensurability with 24\,h makes it difficult to distinguish between 6 or 8\,h periods, 12\,h period possible according to \cite{Dose:21} \\
    33750 & Davehiggins          & 10.5623          & 8.827     & 48    & CE & agrees with the period reported by Sergison listed in the LCDB \\
    40203 & 1998 SP27            & 2.42776           & 5.448     & 23    & C & probably incorrect \\
    43331 & 2000 PS6             & 2.09540            & 7.338     & 33    & E & incorrect \\
    \hline
    \end{tabular}
    \tablefoot{The table lists for each asteroid the period $P_\text{Gaia}$ we derived from Gaia DR3 data, the period $P_\text{LCDB}$ reported in the Lightcurve Database of \cite{War.ea:09}, the number of points $N$ in DR3, the method used for computing periodograms: C -- convex shape models, E -- ellipsoids, CE -- both methods provided the same period. The last column gives our conclusion about the discrepancy between the periods.}
    \tablebib{(1) \url{https://obswww.unige.ch/~behrend/page_cou.html}}
  \end{table*}

  \subsection{Comparison with Gaia DR2 results}
  
  The DR2 data are now part of the DR3; however, the data processing between DR2 and DR3 differs. Therefore, the reported magnitudes can be slightly different. Moreover, DR3 data usually contain additional epochs for each asteroid compared to DR2. Out of 129 models reconstructed from DR2 and published by \citet{Dur.Han:18}, 108 are among our final solutions from DR3. In four cases the rotation periods do not agree within their errors:   asteroids (1540)~Kevola, (2760)~Kacha, (14410) 1991~RR1, and (21904) 1999~VV12. Apart from (2760)~Kacha, the three asteroids were marked as less reliable in \cite{Dur.Han:18} because when 10\% points were removed, the period was not unique. For Kacha, the period of 53\,h derived by an ellipsoidal model from DR2 is two times longer than our new value 26.5\,h derived with a convex model.
  
  \subsection{Comparison with DAMIT}

  There are 1324 asteroids for which we have a model from DR3, and an independent model exists in the Database of Asteroid Models from Inversion Techniques \citep[DAMIT, version from 7 Jun 2022,][]{Dur.ea:10}. For 33 asteroids, their DAMIT periods are different from those we derived from DR3. We list these cases in Table~\ref{tab:inconsistent_DAMIT_periods}. After checking the original data from which DAMIT models were reconstructed, we realized that in most cases, our  DR3 solutions are more reliable, and DAMIT solutions are likely incorrect. These incorrect DAMIT solutions are usually based on sparse data only, often on the Lowell Observatory dataset with poor photometric quality. Only two of our 33 DR3 solutions are clearly wrong; we do not include them in our final dataset. 
  
  
  \begin{table*}
  \tiny
     \caption{Asteroids whose rotation period derived from DR3 was different from that in DAMIT.}
  \label{tab:inconsistent_DAMIT_periods}
  \begin{tabular}{r@{\ }l d d c c c p{7.5cm}}
  \hline
  \multicolumn{2}{c}{Asteroid}  & \multicolumn{1}{c}{$P_\text{Gaia}$}   & \multicolumn{1}{c}{$P_\text{DAMIT}$}  & $N$   & Method    & Ref.         & Comment   \\
                            &   & \multicolumn{1}{c}{[h]}               & \multicolumn{1}{c}{[h]}                &       &           &                   &           \\
  \hline    
  219 & Thusnelda            & 4.44300          & 59.712            & 24  & E           & 1 & incorrect, already in Table~\ref{tab:inconsistent_LCDB_periods} \\
 1040 & Klumpkea             & 37.734            & 56.588            & 33  & CE          & 2 & confirmed by \cite{Pal.ea:20}; DAMIT incorrect \\ 
 1284 & Latvia               & 13.0214           & 9.5506            & 29  & E           & 3 & incorrect \\
 1465 & Autonoma             & 4.88180           & 11.94897          & 44  & C           & 4 & agrees with \cite{Bri:08, Fau.Fau:13, Dit.ea:18}; DAMIT incorrect \\
 1957 & Angara               & 3.67345           & 3.793615          & 37  & CE          & 4  & agrees with \cite{Bin:87}; DAMIT incorrect \\
 2177 & Oliver               & 6.10516           & 6.9969            & 47  & E           & 3  & agrees with the LCDB period; DAMIT incorrect \\
 2180 & Marjaleena           & 7.7030            & 8.34623           & 59  & CE          & 5  & agrees with the LCDB period; DAMIT incorrect \\
 3097 & Tacitus              & 7.4170            & 8.77591           & 50  & CE          & 2 & agrees with \cite{Was.ea:15}; DAMIT incorrect \\
 4846 & Tuthmosis            & 16.3788           & 20.0351           & 46  & CE          & 4  & DAMIT model is likely incorrect \\
 5513 & Yukio                & 5.35953          & 4.81954           & 24  & C           & 3  & DAMIT model is likely incorrect \\
10533 & 1991 PT12            & 8.6007         & 17.20736          & 38    & C         & 6 & DAMIT model is likely incorrect  \\
10579 & Diluca              & 16.4749           & 33.1577           & 63    & C         & 4  & DAMIT model is likely incorrect  \\
11235 & 1999 JP91           & 20.2221           & 20.5136           & 48    & C         & 4  & DAMIT model is likely incorrect \\
11618 & 1996 EX1            & 33.691            & 40.2342           & 37    & CE        & 6 & DAMIT model is likely incorrect \\
12396 & Amyphillips         & 10.4992           & 4.450053          & 53    & CE        & 4  & DAMIT model is likely incorrect \\
12833 & Kamenn\'y \'Ujezd   & 51.232         & 24.731            & 40    & E         & 3  & DAMIT model is likely incorrect \\
13116 & Hortensia           & 47.259            & 37.1773           & 48    & CE        & 4  & confirmed by \cite{Pal.ea:20}; DAMIT incorrect \\
13289 & 1998 QK75           & 438.6             & 43.2618           & 35    & CE        & 4  & DAMIT model is likely incorrect \\
14555 & Shinohara           & 35.650            & 23.856            & 40    & CE        & 4  & DAMIT model is likely incorrect \\
14664 & Vandervelden        & 4.90927           & 2.454632          & 50    & CE        & 4  & DAMIT model is likely incorrect \\
16394 & 1981 QD4            & 3.43229           & 26.8986           & 53    & C         & 4  & DAMIT model is likely incorrect\\
16712 & 1995 SW29           & 17.6578           & 17.17             & 28    & CE        & 6 & DAMIT model is likely incorrect \\
21181 & 1994 EB2            & 82.34         & 72.957            & 32    & C         & 4  & DAMIT model is likely incorrect\\
22018 & 1999 XK105          & 3.54874           & 17.0575           & 27    & CE        & 7 & DAMIT model is likely incorrect \\
27070 & 1998 SA101          & 5.47726           & 4.91448           & 28    & C         & 3  & DAMIT model is likely incorrect \\
29198 & Weathers            & 4.50165        & 4.536367          & 46    & C         & 4  & DAMIT model is likely incorrect \\
32591 & 2001 QV134          & 3.79064           & 3.79346           & 48    & CE        & 4  & only slightly different periods \\
46244 & 2001 HU15           & 5.01382           & 4.877866          & 40    & CE        & 4  & confirmed by \cite{Was.ea:15}; DAMIT incorrect \\
50776 & 2000 FS12           & 11.1477           & 16.7993           & 39    & CE        & 6 & confirmed by \cite{Was.ea:15}; DAMIT incorrect \\
52421 & Daihoji             & 23.191            & 22.5465           & 33    & C         & 6 & DAMIT model is likely incorrect \\
57046 & 2001 KW55           & 6.8831            & 3.418974          & 32    & C         & 6 & no conclusion \\
99667 & 2002 JO1            & 5.6812            & 5.07877           & 26    & C         & 7 & DAMIT model is likely incorrect \\
101537 & 1998 YX14          & 15.7599           & 17.0944           & 25    & C         & 6 & no conclusion \\
    \hline
  \end{tabular}
  \tablefoot{The table lists for each asteroid the period $P_\text{Gaia}$ we derived from Gaia DR3 data, the period $P_\text{DAMIT}$ from the DAMIT database, the number of points $N$ in DR3, the method used for computing periodograms: C -- convex shape models, E -- ellipsoids, CE -- both methods provided the same period, and the reference to the DAMIT model. The last column gives our conclusion about the discrepancy between the models.}
  \tablebib{
    (1)~\citet{Mar.ea:21}; (2)~\citet{Han.ea:13b}; (3)~\citet{Dur.ea:20}; (4)~\citet{Dur.ea:19}; (5)~\citet{Han.ea:18a}; (6)~\cite{Dur.ea:18c}; (7)~\cite{Dur.ea:16}.
}
  \end{table*}

  \subsection{Final models}
  \label{sec:final_models}
  
  Out of the 8602 asteroids that passed all the checks (Sect.~\ref{sec:stable_solutions}), 6 were excluded based on the period comparison with LCDB and DAMIT (Tables~\ref{tab:inconsistent_LCDB_periods} and \ref{tab:inconsistent_DAMIT_periods}), and so there are 8596 final models. Their spin solutions are available at the CDS; the first and the last part of the table are shown in Table~\ref{tab:results} as an example. New models that are not in DAMIT will be uploaded there. The false DAMIT models identified in Table~\ref{tab:inconsistent_DAMIT_periods} will be updated.

  As expected, the success rate of deriving a unique spin solution and a corresponding shape model depends on the number of observations. In Fig.~\ref{fig:histogram} the orange histogram shows the distribution of the number of observations $N$ in our final sample of models. The plot also shows the success rate as the ratio of the number of models we derived to the total number of asteroids in the DR3 sample for each bin. It starts at a few percent for $N = 21$ and increases to about 40\% for $N$ around 60. For even larger $N$, the number of asteroids per bin is small, and therefore the ratio fluctuates significantly.
  
  \begin{table*}
  \tiny
    \caption{Spins of asteroids derived from DR3 photometry.}
    \label{tab:results}
    \begin{tabular}{r@{\ }l r r r r d r c}
    \hline
    \multicolumn{2}{c}{Asteroid}    & \multicolumn{1}{c}{$\lambda_1$} & \multicolumn{1}{c}{$\beta_1$}   & \multicolumn{1}{c}{$\lambda_2$} & \multicolumn{1}{c}{$\beta_2$}   & \multicolumn{1}{c}{$P$}   & \multicolumn{1}{c}{$N$}   & Method    \\
                            &       & \multicolumn{1}{c}{[deg]}       & \multicolumn{1}{c}{[deg]}       & \multicolumn{1}{c}{[deg]}       & \multicolumn{1}{c}{[deg]}       & \multicolumn{1}{c}{[h]}   &                           &           \\
    \hline    
      5 & Astraea                   &   122 &    37 &   312 &    40 &  16.8008 &    36 &   C  \\
     26 & Proserpina                &    83 & $-46$ &   255 & $-56$ &  13.1092 &    58 &   C  \\
     32 & Pomona                    &   103 &    39 &   279 &    41 &   9.4475 &    38 &  CE  \\
     33 & Polyhymnia                &    20 & $-29$ &   197 & $-31$ &  18.6088 &    36 &   C  \\
     41 & Daphne                    &   207 & $-35$ &   355 & $-39$ &  5.98807 &    35 &   C  \\
     43 & Ariadne                   &    69 & $-10$ &   252 & $-12$ &  5.76182 &    53 &   C  \\
     44 & Nysa                      &    96 &    45 &   287 &    54 &   6.4214 &    25 &  CE  \\
     45 & Eugenia                   &   123 & $-32$ &   299 & $-20$ &  5.69918 &    30 &   C  \\
     48 & Doris                     &   107 &    32 &   291 &    45 &  11.8900 &    50 &   C  \\
     50 & Virginia                  &   103 &    12 &   281 &    26 &  14.3107 &    31 &   E  \\

    \vdots &                 &       &       &       &       &           &       &   \\  
 301892 & 1998 QL98                 &    97 & $-63$ &       &       &  3.81273 &    28 &   C  \\
 307960 & 2004 GS75                 &    23 &    71 &       &       &  12.7888 &    30 &  CE  \\
 313568 & 2003 DW4                  &    96 &    63 &       &       &  4.53729 &    25 &   E  \\
 329036 & 2011 AE35                 &    18 & $-50$ &       &       &   29.959 &    22 &   E  \\
 334190 & 2001 SQ199                &    22 & $-61$ &   162 & $-41$ &   7.3653 &    30 &   C  \\
 353971 & 2000 AE210                &     9 & $-50$ &   275 & $-80$ &   21.467 &    26 &   E  \\
 354510 & 2004 PV66                 &    50 & $-54$ &       &       &   6.6110 &    29 &   E  \\
 362935 & 2012 JB5                  &    89 &    53 &       &       &  12.6190 &    68 &   C  \\
 380282 & 2002 AO148                &   330 &    81 &       &       &  10.6120 &    29 &   C  \\
 397797 & 2008 MZ2                  &    91 &    76 &       &       &   49.936 &    35 &  CE  \\
    \hline
  \end{tabular}
  \tablefoot{The table lists the first ten  and the last ten asteroid models derived from Gaia DR3 data. The full table is available in  electronic format at CDS. For each asteroid we report its spin axis direction in ecliptic latitude $\lambda_1$ and longitude $\beta_1$ (the second pole solution has ecliptic coordinates $\lambda_2$ and longitude $\beta_2$), the sidereal rotation period $P$, the number $N$ of photometric measurements in DR3, and the method used for computing periodograms: C -- convex shape models, E -- ellipsoids, CE -- both methods provided the same period. The uncertainty of the rotation period $P$ is on the order of the last decimal place.}
  \end{table*}

  \subsection{Uncertainty of shape models}
  \label{sec:shape_stability}

\begin{figure}
\begin{center}
\small
20 points \quad pole $(247^\circ, -15^\circ)$ \quad resolution 3, 3
\includegraphics[width=\columnwidth, trim=1.5cm 7cm 1cm 1cm, clip]{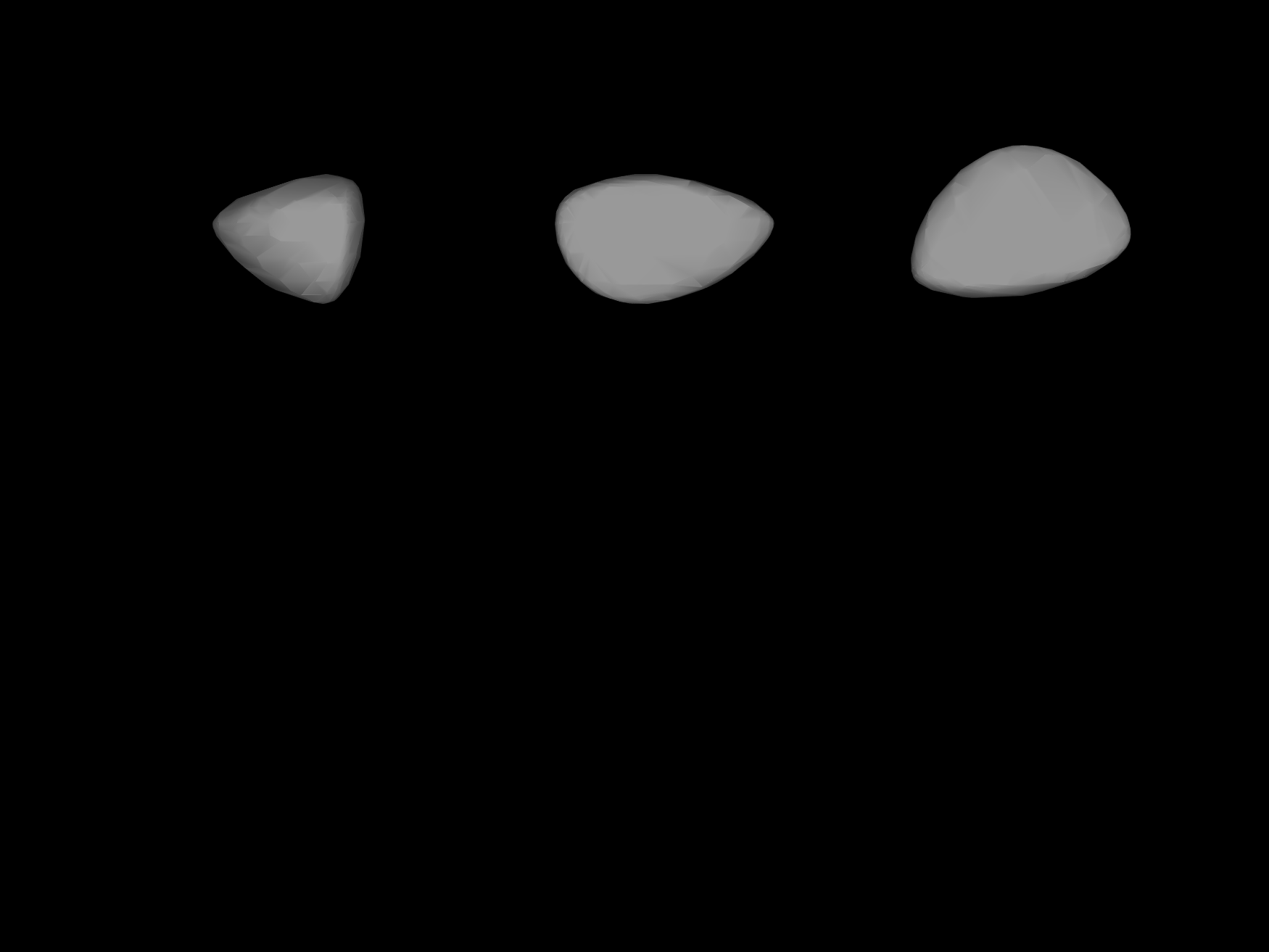}\\
30 points \quad pole $(251^\circ, -14^\circ)$ \quad resolution 3, 3
\includegraphics[width=\columnwidth, trim=1.5cm 7cm 1cm 1cm, clip]{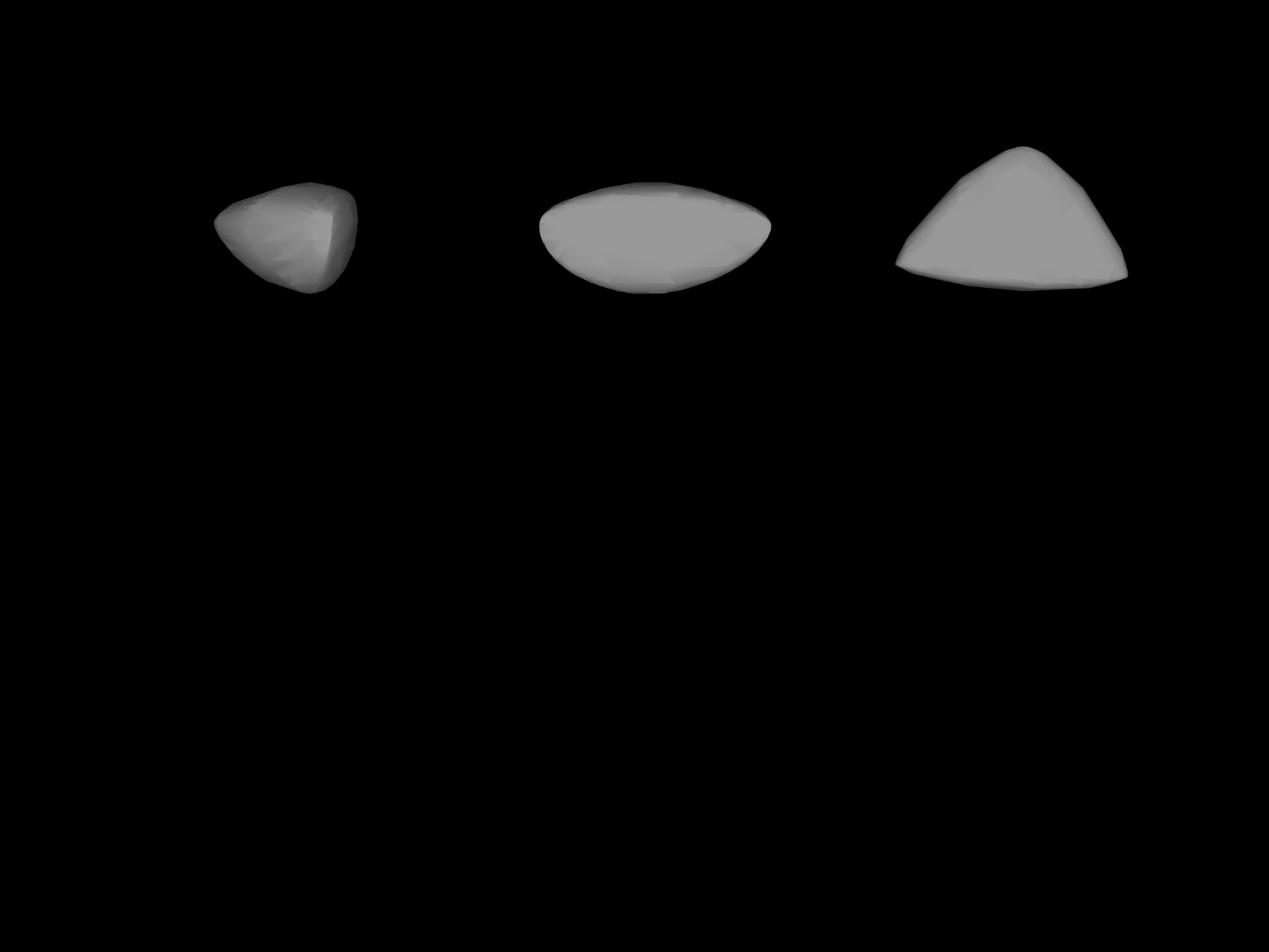}\\
40 points \quad pole $(250^\circ, -15^\circ)$ \quad resolution 3, 3
\includegraphics[width=\columnwidth, trim=1.5cm 7cm 1cm 1cm, clip]{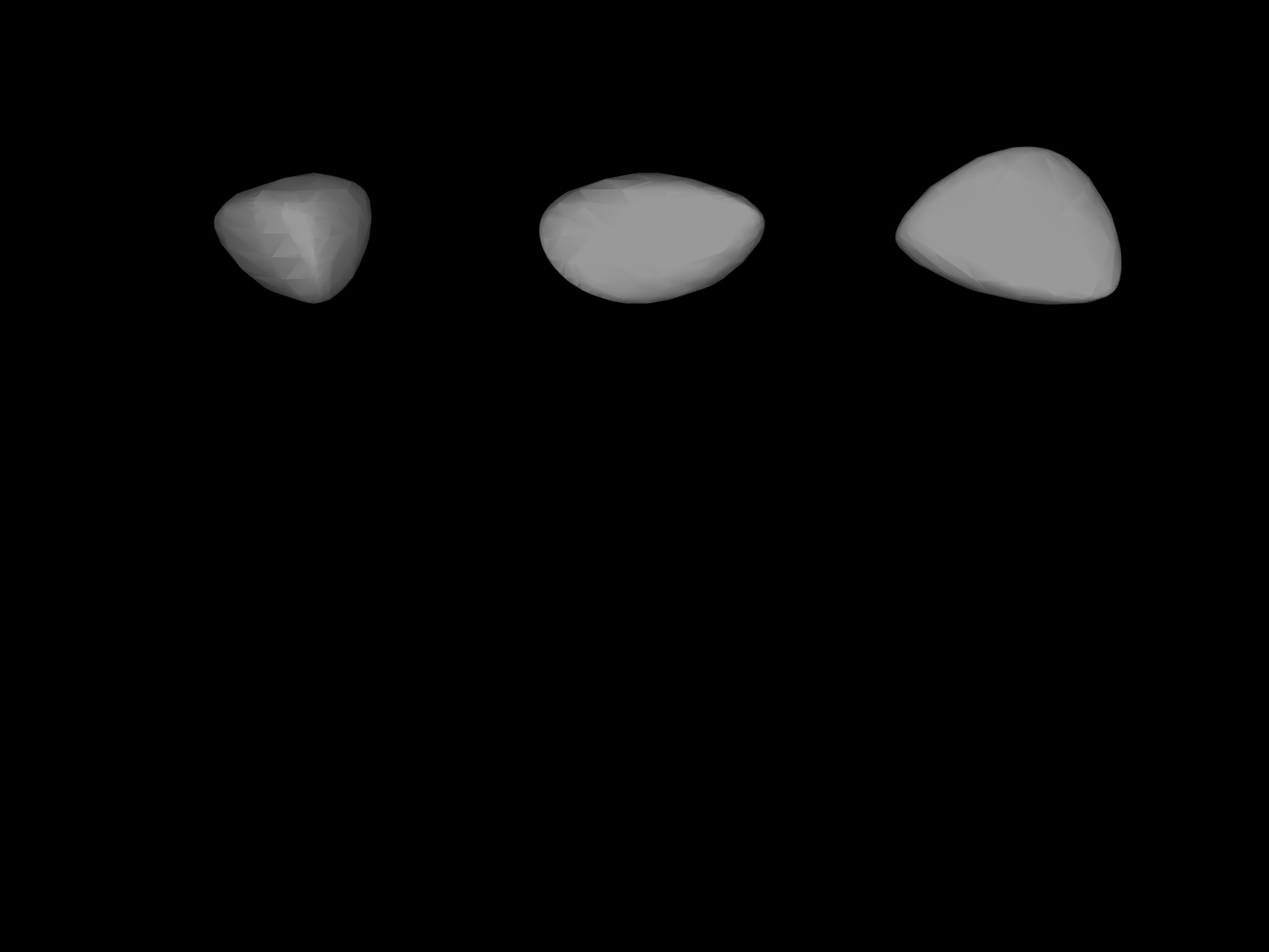}\\
53 points \quad pole $(253^\circ, -15^\circ)$ \quad resolution 3, 3
\includegraphics[width=\columnwidth, trim=1.5cm 7cm 1cm 1cm, clip]{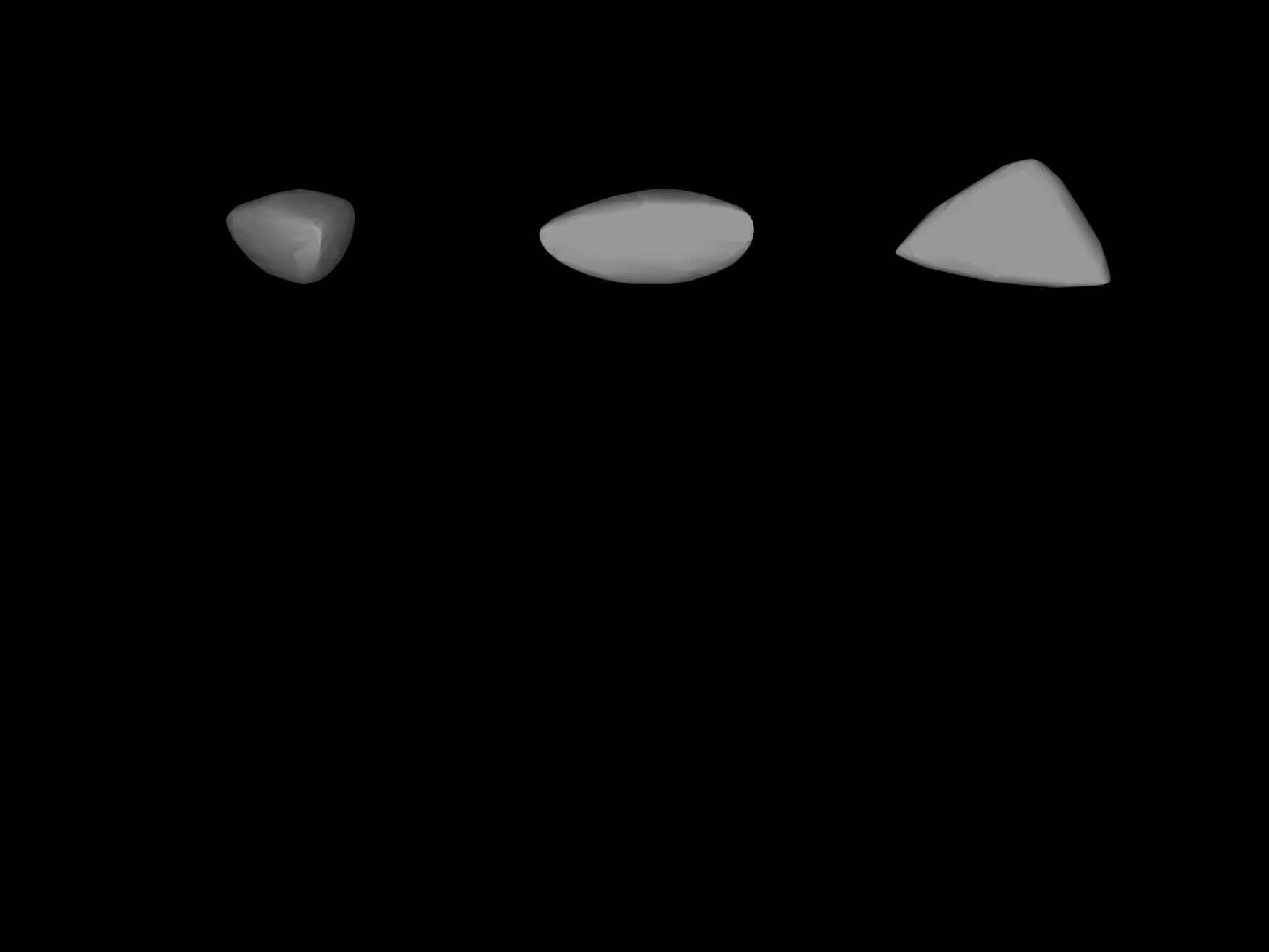}\\
53 points \quad pole $(252^\circ, -15^\circ)$ \quad resolution 4, 4
\includegraphics[width=\columnwidth, trim=1.5cm 7cm 1cm 1cm, clip]{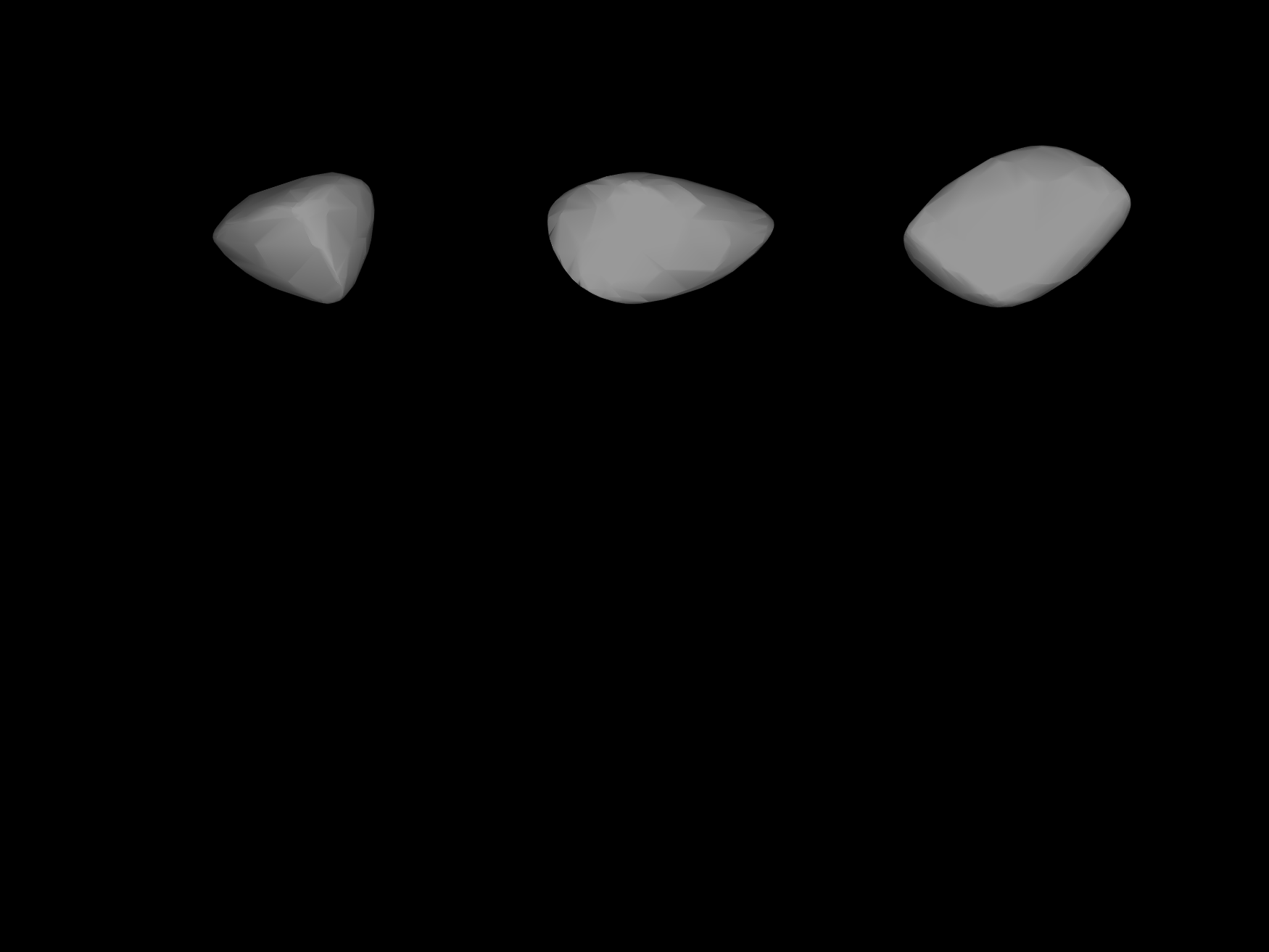}\\
53 points \quad pole $(252^\circ, -12^\circ)$ \quad resolution 5, 5
\includegraphics[width=\columnwidth, trim=1.5cm 7cm 1cm 1cm, clip]{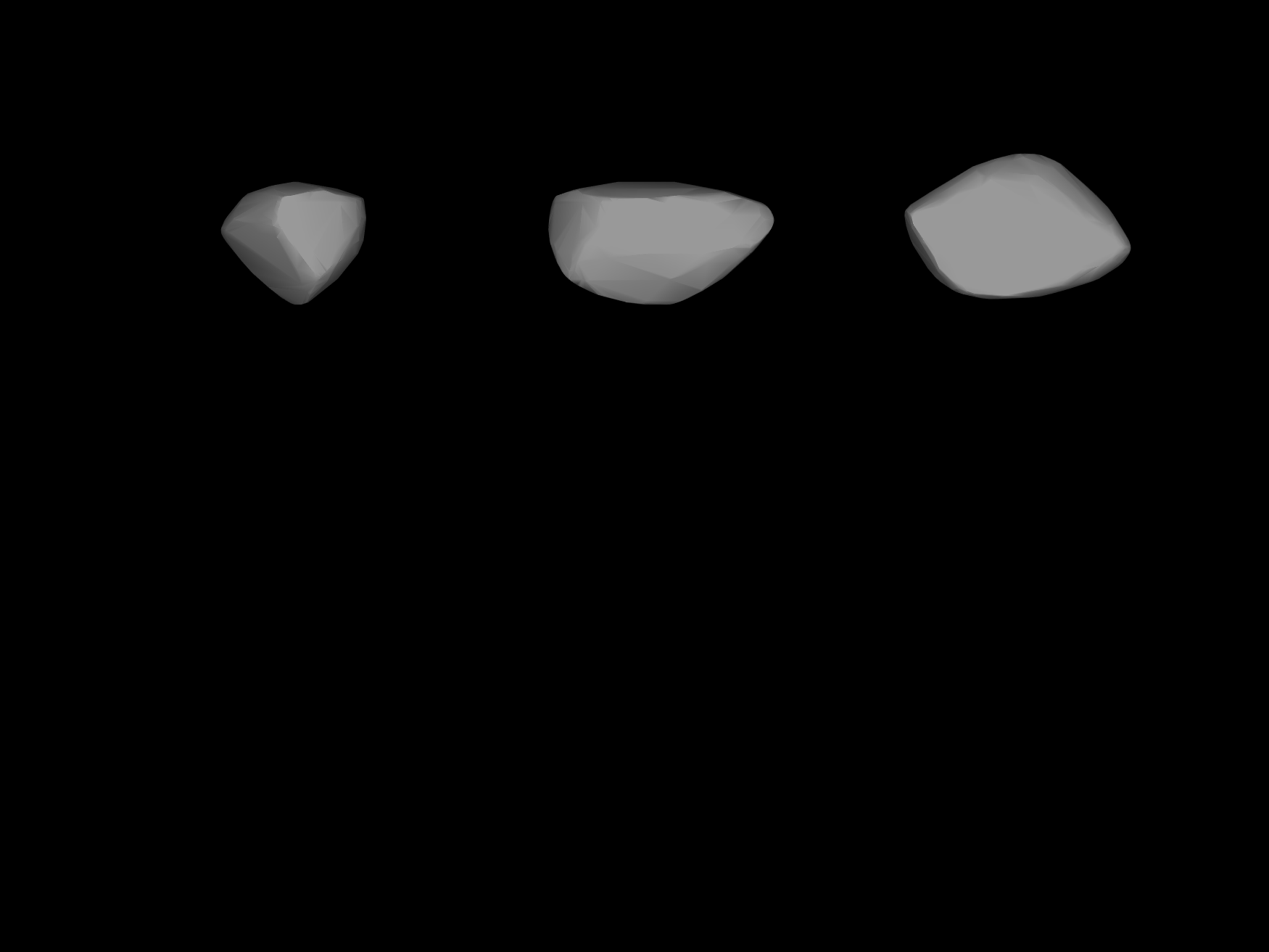}\\
\end{center}
\caption{\label{fig:shape_stability} Shape models of (43) Ariadne reconstructed from different numbers of data points and with different degrees and orders of spherical harmonics series denoted as resolution.}
\end{figure}

  Compared to ground-based surveys, the accuracy of Gaia asteroid photometry is much higher. This high photometric accuracy enables us to reconstruct asteroid rotation states reliably from fewer measurements than we would need with ground-based data. With Gaia, we need tens of data points, while with less accurate photometry we would need hundreds of them \citep{Dur.ea:05, Dur.ea:20}. However, with only tens of observations the reconstructed shape model is sensitive to the number of observations, their distribution in time, and the model parameterization. We demonstrate this in Fig.~\ref{fig:shape_stability}, where we show shape models of asteroid (43)~Ariadne reconstructed from a different number of observations going from 20 to 53 (the full DR3 set). The shape is also sensitive to the model resolution described by the order and degree of spherical harmonics series that describe the surface curvature \citep{Kaa.Tor:01}.

  While spins derived from DR3 are reliable and stable with respect to the perturbations of input data (e.g., the dispersion in spin directions for different shape models in Fig.~\ref{fig:shape_stability} is $\pm 2^\circ$ in ecliptic latitude $\beta$ and $\pm 3^\circ$ in ecliptic longitude $\lambda$), the shape models are not, and any results based on shape parameters should take this into account.
  We do not report uncertainties of spin parameters in Table~\ref{tab:results}. We expect them to be smaller than $20^\circ$, which was a mean difference between pole directions of models derived from Gaia DR2 and independent DAMIT models \citep{Dur.Han:18}. Uncertainty in the sidereal period $P$ is typically a fraction ($\sim 1/10$) of the distance between local minima $\Delta P$.

  \section{Results}\label{sec:results}

    In this section we analyze the spin and shape properties of asteroids we reconstructed from Gaia photometry (Sect.~\ref{sec:method}). We also search for correlations with other physical properties adopted from the literature. This includes diameters and family classification from the MP3C\footnote{\url{https://mp3c.oca.eu/}} database, and family ages from \cite{Nesvorny2015}.

    Due to the symmetry of the inverse problem \citep{Kaa.Lam:06}, we usually have two pole solutions with similar pole latitudes and differences in longitudes of $\sim$180$^\circ$. Their order in Table~\ref{tab:results} is given by the quality of the fit, so always selecting the first one should not bias the analysis of physical properties. In some cases we plot both solutions in the figures. The poles are expressed in the ecliptic coordinate frame ($\lambda$, $\beta$). Computing the pole obliquity $\varepsilon$ for each model (the angle between the spin vector and the normal to the orbital plane) requires a standard transformation into the orbital reference frame.

    Due to the low number of optical measurements, the shape is only loosely constrained (Sect.~\ref{sec:shape_stability}). So instead of using the whole shape information, we utilize only the elongation $a/b$, which is determined as the axis ratio of the dynamically equivalent ellipsoid. It is the primary and most stable parameter of the shape. 
  
  \subsection{Spin distribution in the main belt and beyond}

\begin{figure*}
\begin{center}
\resizebox{1.0\hsize}{!}{\includegraphics{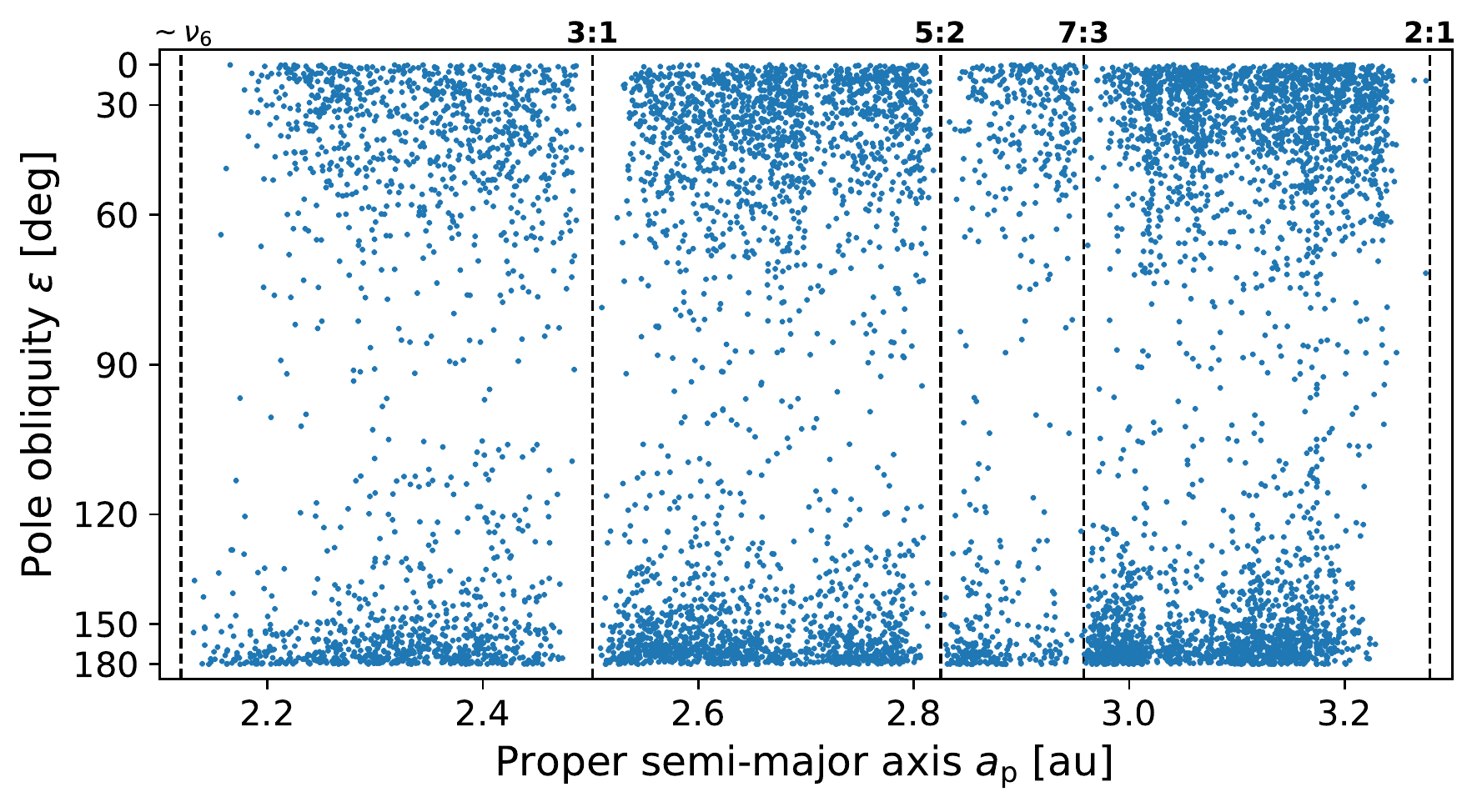}\includegraphics{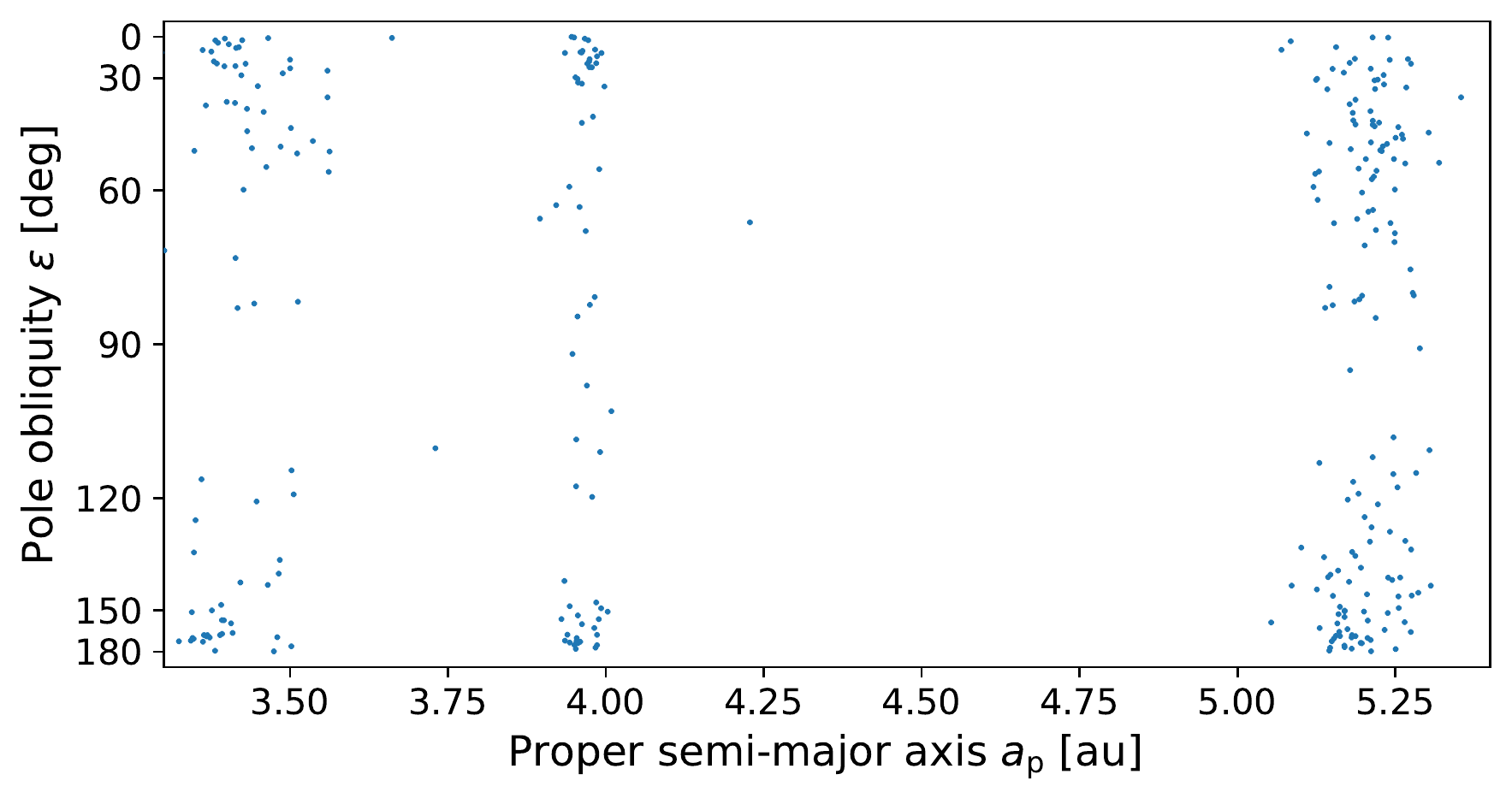}}\\
\end{center}
\caption{\label{fig:ab1} Distribution of all models in the proper  semimajor axis and spin obliquity in the asteroid main belt ({\it left\/}) and for asteroids with $a>3.3$\,au:  Cybeles, Hildas, and Jovian Trojans ({\it right\/}).}
\end{figure*}

\begin{figure}
\begin{center}
\resizebox{1.0\hsize}{!}{\includegraphics{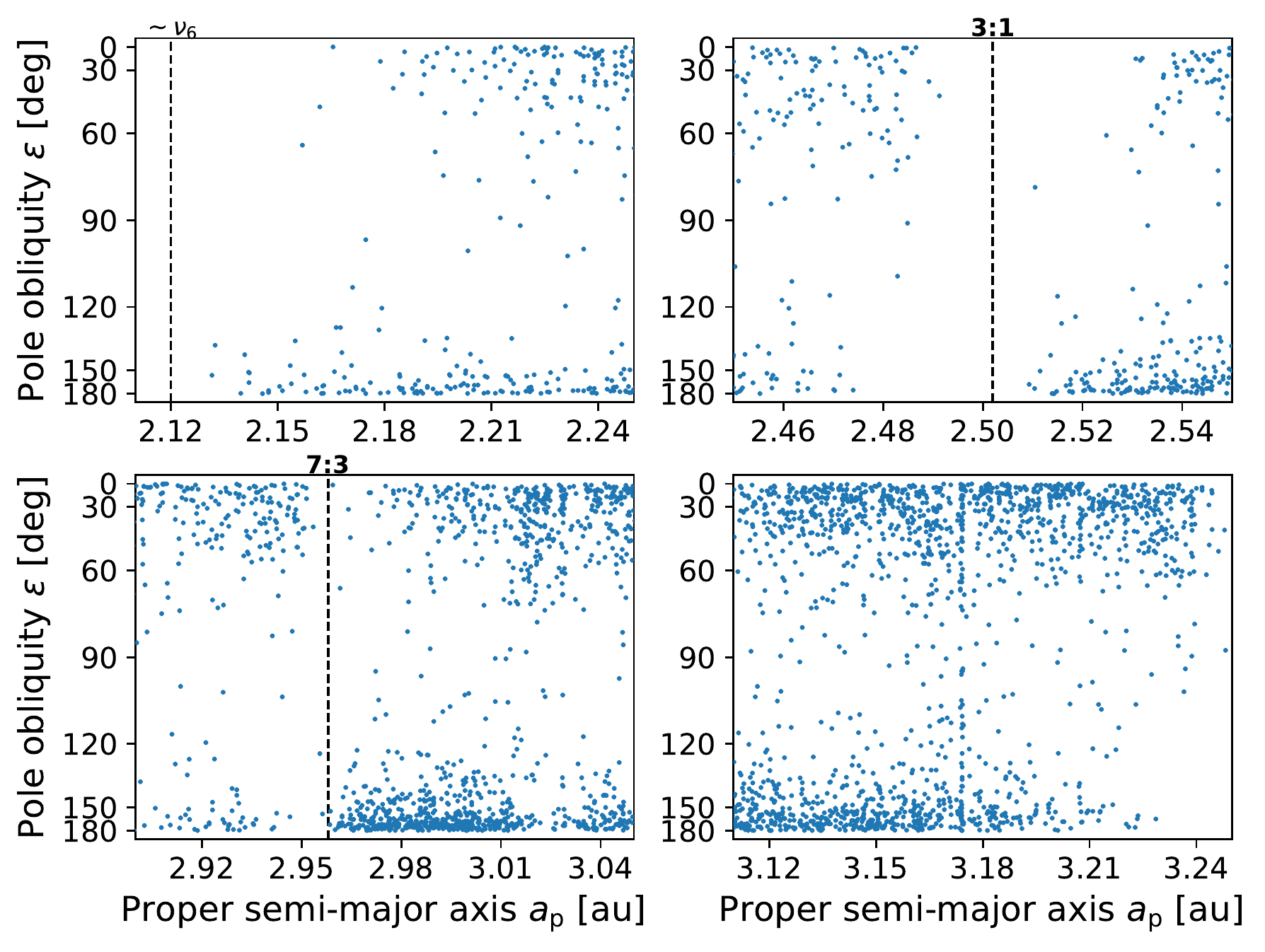}}\\
\end{center}
\caption{\label{fig:ab2} Distribution of models in the proper semimajor axis and spin obliquity in several regions of the main belt. Top left panel: Near $\nu_6$ resonance. Top right panel: Along 3:1 resonance. Bottom left panel: Along 7:3 resonance. Bottom right panel: Outer main belt.}
\end{figure}

    The distribution of pole obliquities in the main belt is shown in Fig.~\ref{fig:ab1}. For visualization purposes, the scale of the vertical axis is linear in $\cos\varepsilon$, so an isotropic distribution of spins would have a uniform distribution in $\cos\varepsilon$. We also use this scale   in other figures when plotting the distribution of obliquity $\varepsilon$. As expected \citep[see, e.g.,][]{Hanus2011}, the distribution is far from uniform.
    Poles are clustered toward extreme obliquities with fewer asteroids with poles close to their orbital plane ($\varepsilon\sim90^\circ$). Although there is a selection bias against asteroids with poles close to the ecliptic plane, it cannot be responsible for the observed distribution. Moreover, the distribution also depends on the asteroids' size. This has been interpreted as the evolution caused by the Yarkovsky--O'Keefe--Radzievskii--Paddack (YORP) effect. The plot also nicely demonstrates the evolution due to the Yarkovsky effect, which causes the retrograde asteroids ($\varepsilon > 90^\circ$)  to migrate inward  to the smaller semimajor axis, while prograde rotators ($\varepsilon < 90^\circ$) migrate outward. This causes regions close to mean-motion resonances with Jupiter to be depleted;  prograde rotators lean to the resonance from the left, get scattered when crossing it, and the space on the right does not contain prograde rotators. The situation is symmetric for retrograde rotators. We illustrate this coupling in more detail near the $\nu_6$, 3:1 and 7:3, and the outer edge of the main belt in Fig.~\ref{fig:ab2}. The concentration of asteroids near $a_\mathrm{p}\sim3.174$\,au is caused by the 5-2-2 three-body resonance causing the chaotic diffusion in the Veritas collisional family \citep{Tsiganis2007}. This group has a smaller fraction of extreme obliquities compared to the main belt. As the Veritas family is quite young \citep[$\sim8$~Myr,][]{Carruba2017}, the spin vectors of its members are likely less evolved by the YORP.
    
    Another fact that is demonstrated on the plots is that the prograde--retrograde distribution is not symmetric;  spins of retrograde models are more tightly clustered toward the perpendicular orientation than those of prograde models. This might be caused by secular spin-orbital resonances, which primarily affect only the prograde rotators. However, we do not provide any further analysis here.

    The asymmetry in positions of prograde and retrograde rotators is present, perhaps to a lesser extent, also in the Cybeles, Hildas, and Jupiter Trojans groups. However, we note two possible caveats here. These groups, on average, contain larger bodies than main-belt asteroids, due to their larger heliocentric distance and dark albedos (i.e., comparable S/N of the DR3 fluxes are for different sizes). In addition, due to the short time span of only about three years of DR3 data, there might be some observational bias correlated with the semimajor axis because the viewing geometry has changed less than for main-belt asteroids. For this reason, spin and shape determination for distant asteroids could be less reliable.
  
  \subsection{Distribution of rotation periods}

\begin{figure}
\begin{center}
\resizebox{1.0\hsize}{!}{\includegraphics{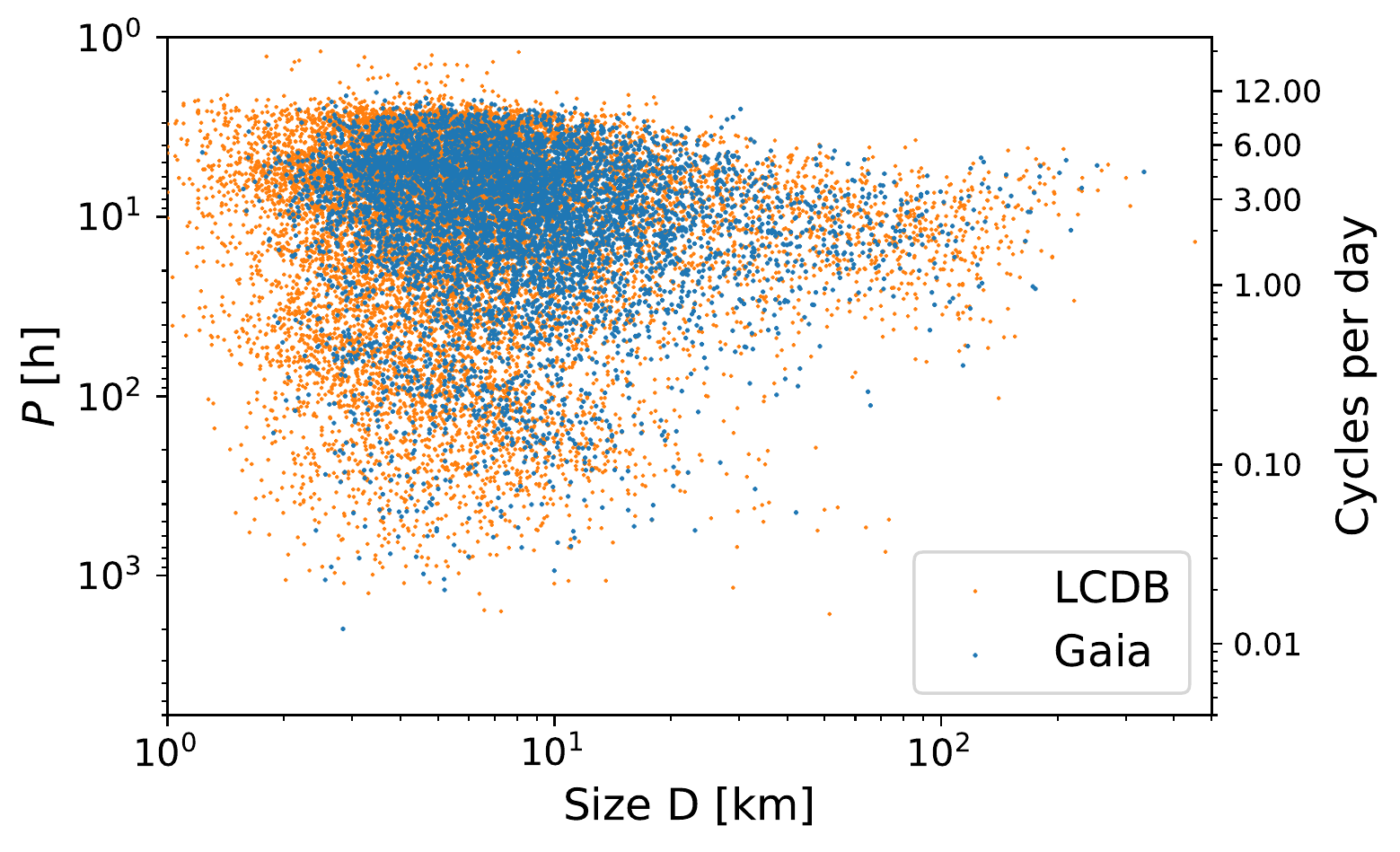}}\\
\end{center}
\caption{\label{fig:dp}Rotation period $P$ as a function of asteroid size $D$. We plot the solution based on DR3 data (blue) and reliable periods (with validity flag U=3 or 2) from the LCDB database (orange).}
\end{figure}

\begin{figure*}
\begin{center}
\resizebox{1.0\hsize}{!}{\includegraphics{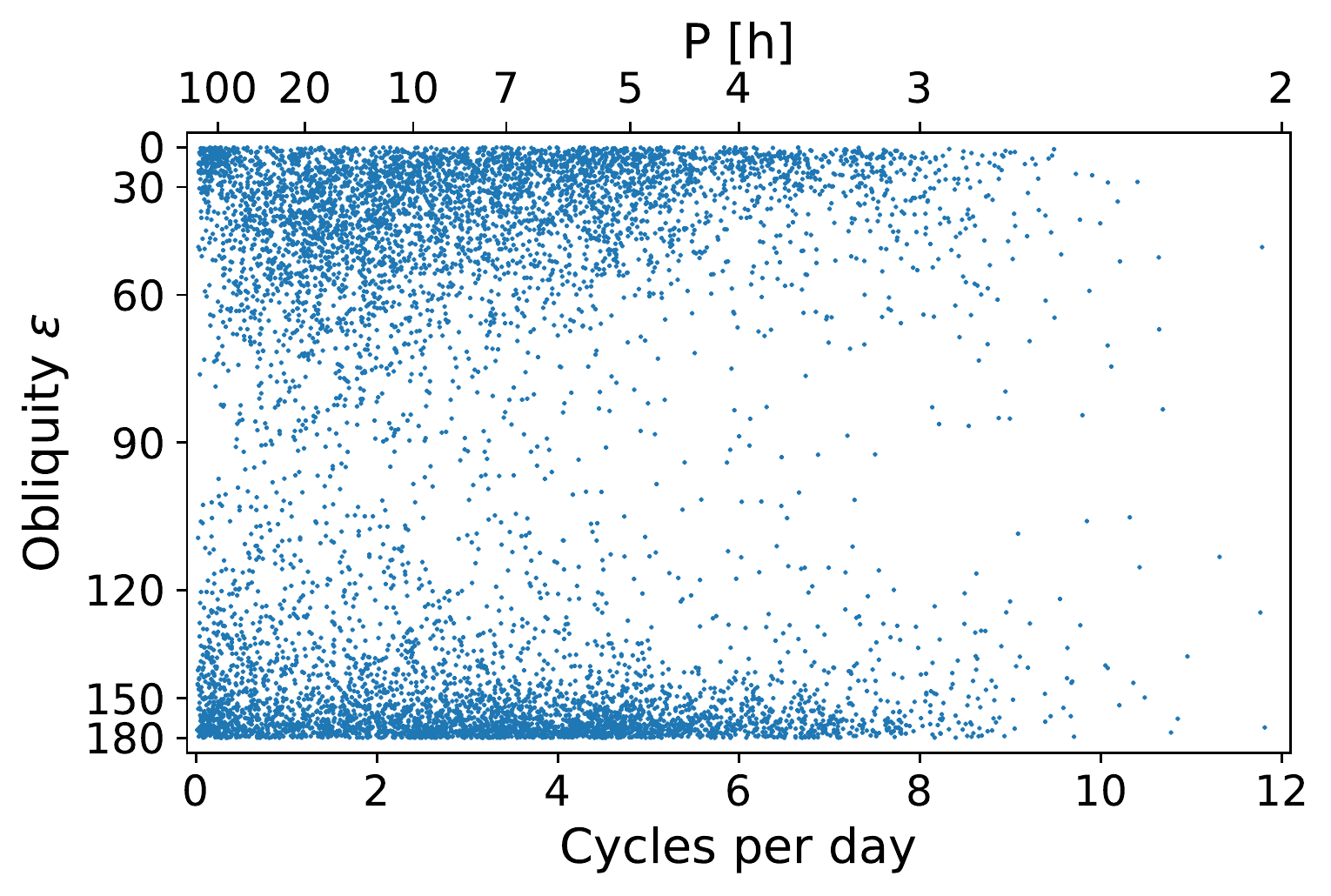}\includegraphics{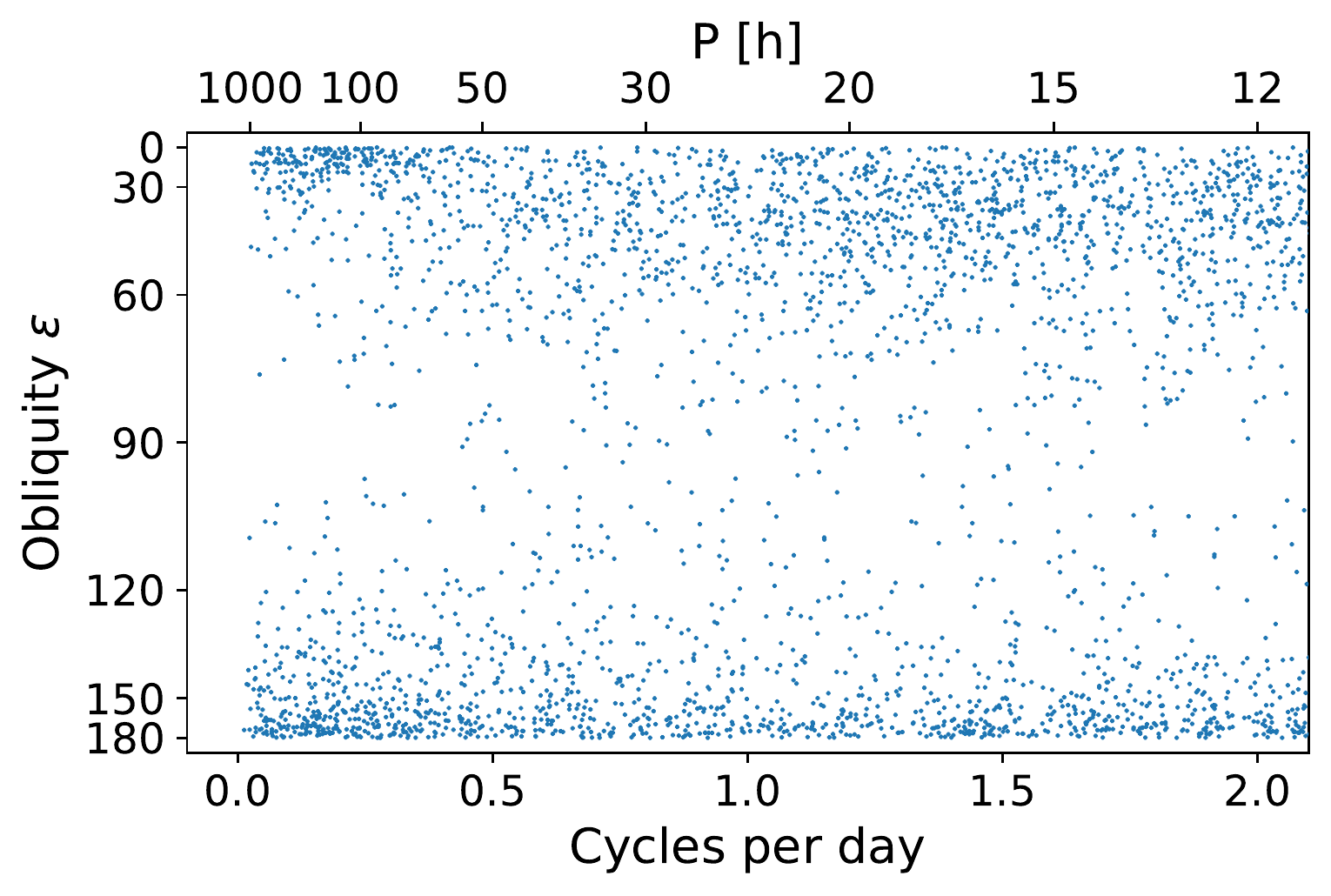}}\\
\end{center}
\caption{\label{fig:perBeta}Rotation periods vs. pole-obliquity $\varepsilon$ for Gaia DR3 models. The left panel shows all asteroids, the right panel shows asteroids with periods $P\gtrsim12$\,h.}
\end{figure*}

    In Fig.~\ref{fig:dp} we show the distribution of rotation periods with respect to the size and also include the reliable solutions from the LCDB (flag U = 3 or 2). Clearly, the two samples (i.e., Gaia and LCDB) are different due to the observational bias:  Gaia lacks larger bodies as those were often saturated, but also small kilometer and subkilometer-sized objects that were either faint for Gaia or did not have frequent close approaches. The small asteroids in LCDB are almost exclusively NEAs that have rare close approaches, but if they do, they can get bright enough even for small aperture telescopes. Moreover, it is relatively simple to obtain a reliable period during the close approach, but to derive a spin state and shape we need more approaches with different observing geometries. We do not have such data from DR3.  DR3 data also allowed us to derive many solutions with rotation periods $>50$\,h. Not many such solutions are available in the LCDB, due to the observational bias, because long periods are challenging for ground-based observatories that produce dense light curves \citep{Mar.ea:15}.

    The most intriguing feature in the distribution of rotation periods is the group of slower rotators ($P\gtrsim50$\,h) separated by a gap from the faster rotators. A similar behavior was observed for the  Jovian Trojans \citep{Kalup2021}. In the absence of the YORP effect, the period bimodal distribution was thought to be caused by the presence of slowly rotating bodies that are believed to be originally synchronized equal-sized binaries. This scenario is unlikely to explain the period bimodality in the population of the small main-belt asteroids. Instead, rotation periods of smaller asteroids ($D<30$\,km)  are evolved by YORP;  they gradually increase or decrease, sometimes even changing the sign of the evolution due to possible changes in surface topography (i.e., after a collision, due to close approaches with terrestrial planets, or after landslides due to fast rotation). If the period decreases, the asteroid can even start to rotate in the non-principle axis state (known as tumbling) as its rotation energy gets low enough to become excited even by a minor impact. However, this scenario predicts that the YORP-induced change in the period is smooth; there should not be any gap between the faster and slower rotators. In the group of slower rotators, their rotation period correlates with their size:  larger bodies have longer rotation periods. 
    
    In Fig.~\ref{fig:perBeta} we illustrate the dependence of the pole obliquity on the rotation period. Bodies with periods $P<50$\,h are qualitatively similar to the general trends in the population;    prograde and retrograde rotators both tend to have obliquities close to the YORP end states of $\epsilon\sim0$ or $\sim180$ degrees, while prograde rotators exhibit a larger scatter. However, the slower rotators ($P<50$\,h) violate this behavior. The range of obliquities is larger for retrograde rotators and smaller for prograde ones. The behavior for prograde rotators with $P > 100$\,h is particularly puzzling as basically all solutions have $0<\varepsilon<30^\circ$. 
 
    We note a possible bias in our dataset concerning the slower rotators. Many tumblers were found in this population, and we should have many in our dataset as well. However, the convex inversion model we used assumes a relaxed rotation state and is inadequate for tumblers. This could be indicated by worse fits to the data;  the average RMS of our solutions should be larger for the tumblers compared to the faster-rotating bodies. We observe larger  RMS values for bodies with $P>50$\,h (see Fig.~\ref{fig:rms}),  so it is likely that some asteroids for which we derived a model are not exactly in the state of principal-axis rotation. However,  they are only slightly excited so that their sparse photometry can still be modeled with a single-period model.
  
    The DR3 shape models exhibit a dependency between the size $D$ and the shape elongation $a/b$ (see Fig.~\ref{fig:dab}), which is in good agreement with previous reports (e.g.,  \citealt{Cibulkova2016, Szabo2022}). Asteroids larger than $\sim30$\,km are less elongated (i.e., more spherical) than the smaller ones.
 
    Finally, Fig.~\ref{fig:dab} illustrates the shape elongation dependence on the asteroid's rotation period. Highly elongated asteroids with $a/b \gtrsim 1.5$ are rare for rotation periods shorter than about 3\,h. In general, the elongation increases with increasing rotation period with a peak near 6\,h. The elongation then slowly decreases for periods in an interval of $\sim6$--50\,h. After that, the elongation increases and has the largest mean values for this population of rotators. Our results agree with \citet{Szabo2022}, who based their study on almost 10,000 rotation periods derived from TESS data. Contrary to our results, \citet{Szabo2022} did not report the increase in  the mean elongation for asteroids with $P>50$\,h that we see in the DR3 models. However, it is unclear whether this is a real effect or a systematic bias related to the  higher RMS values discussed in the previous paragraph.

  \subsection{Spin vector distribution in asteroid families}
  
    The Yarkovsky-driven evolution is also seen in asteroid families, where the $1/D$ dependence of the Yarkovsky drift on the diameter $D$ causes the typical V-shaped spreading in the proper semimajor axis. In Figs.~\ref{fig:fam1},~\ref{fig:fam2},~and~\ref{fig:fam3}, we show 14 asteroid families with the highest number of asteroid models. In general, the distribution of prograde--retrograde models agrees with the theoretical expectations that prograde rotators are to the right (larger $a$) from the center of the family, while retrograde ones are on the left (smaller $a$) (see the  plots for the Eos, Eunomia, Themis, and Koronis families). However, this trend is not as evident in some other families (e.g.,  Dora and Euphrosyne). Families that are truncated by resonances can consist of just one wing to which correspond either prograde or retrograde rotators. Typical examples of truncated families are Phocaea and Maria, both having ``well-behaved'' spin directions.

    There is no family where all bodies follow the ideal Yarkovsky-driven evolution. Specifically, bodies close to the family center or at its outskirts often have spin that is inconsistent with the expected Yarkovsky drift. The former is likely due to the slow evolution of some bodies, especially if their initial locations in semimajor axis $a$ were the most extreme,  on the opposite side with respect to their sense of rotation. Bodies at the borders of the family are then likely interlopers mistakenly associated with the family by the HCM method \citep{Zappala1990}. In general, interlopers can be present at any place in the family, but their number should not be larger than $\sim$10\%. Moreover, non-catastrophic collisions can randomize the spin states of the family members at any location. The importance of these collisions is given by the collisional timescale and the family age. These timescales are dependent on the asteroid size and are usually  on the same order as the age of the family for bodies with sizes of $\sim$10--30\,km.

    So far, we do not see any clear sign of the stochastic YORP evolution \citep{Sta:09,Bot.ea:15} that should  cause the spin orientation of small family members to be oriented randomly, nor following the prograde--retrograde dichotomy of larger bodies.
  
  \subsection{Rotation periods in families}
 
    The distribution of rotation periods in several asteroid families was recently studied by \citet{Szabo2022}. The authors report that rotation period distributions differ between the cores and outskirts  of some collisional families (e.g., Flora and Maria) while are consistent among some other families. Moreover, the authors also show that the lightcurve amplitude distributions in families could be correlated with the family age, probably due to the temporal evolution of asteroid shapes. Here we focus on the same correlations based on the properties derived from DR3 photometry. Figure~\ref{fig:cumPerFam1} shows the distribution of rotation periods and elongations in various families. We ordered the families according to their approximate age \citep[parameter $c_0$ from][]{Nesvorny2015}. For the distribution of the rotation periods, we selected only asteroids with $P<24$\,h to remove the population of slower rotators. We see a clear trend in Fig.~\ref{fig:cumPerFam1}: older families have a larger fraction of asteroids with $P=10$--24\,h and with rounder shapes (smaller $a/b$), which likely indicates a more evolved population in the period due to YORP spin down. 
  
  \section{Conclusions}\label{sec:conclusions}

    Although  the DR2 already showed the scientific potential of Gaia asteroid photometry \citep{Dur.Han:18, Mom.ea:18, Col.ea:21, Wil.ea:22}, it is the DR3 that enables us to make a significant leap forward in asteroid modeling. We derived $\sim$8600 models from Gaia DR3, which is more than a factor of two higher than what is currently available in the DAMIT database ($\sim$3500). Considering the overlap between our models from DR3 and those already in DAMIT (about 1300 asteroids), we obtained $\sim$7300 new shape and spin state models. So  we now have information about   spin axis direction, rotation period, and shape  for more than  ten thousand asteroids.

    The analysis of the distribution of asteroid spins we present in this paper confirms previous findings and the expected trends. Specifically, the spins are affected by the YORP effect:  they evolve toward extreme obliquity values. Some asymmetry between prograde and retrograde rotators might be related to spin-orbital resonances. Prograde and retrograde rotators have opposite Yarkovsky drifts on the semimajor axis, which leads to the separation of these two groups near the prominent mean-motion resonances and in asteroid families. We also see correlations between obliquity, rotation period, and shape elongation. 

    The results presented in this paper are based solely on Gaia DR3 data. The number of new asteroid models is large compared to the number of models currently known, but still small compared to the number of asteroids for which DR3 photometry is available. The next step is to combine DR3 with photometry from other surveys. This will increase the number of data points for individual asteroids, enlarge the time span of observations, and eventually lead to thousands of additional asteroid models.

    In future Gaia data releases, the number of asteroids will not increase dramatically. However, the number of observations per object increases as Gaia continues to collect data, so with a three times longer observing window, for example, most asteroids will have more than 30 detections (see Fig.~\ref{fig:histogram}), and we expect that it will be possible to reconstruct spin states for tens of thousands of asteroids. With more data points, the uncertainty of spin parameters and shape will decrease.
  
  \begin{acknowledgements}
    The authors were supported by the grant 20-08218S of the Czech Science Foundation and used the computational cluster Chimera of the Faculty of Mathematics and Physics, Charles University.
    This work has made use of data from the European Space Agency (ESA) mission {\it Gaia} (\url{https://www.cosmos.esa.int/gaia}), processed by the {\it Gaia} Data Processing and Analysis Consortium (DPAC, \url{https://www.cosmos.esa.int/web/gaia/dpac/consortium}). Funding for the DPAC has been provided by national institutions, in particular, the institutions participating in the {\it Gaia} Multilateral Agreement.
  \end{acknowledgements}

  \newcommand{\SortNoop}[1]{}

\begin{appendix}
\section{Supplementary figures}

\begin{figure*}
\begin{center}
\resizebox{0.5\hsize}{!}{\includegraphics{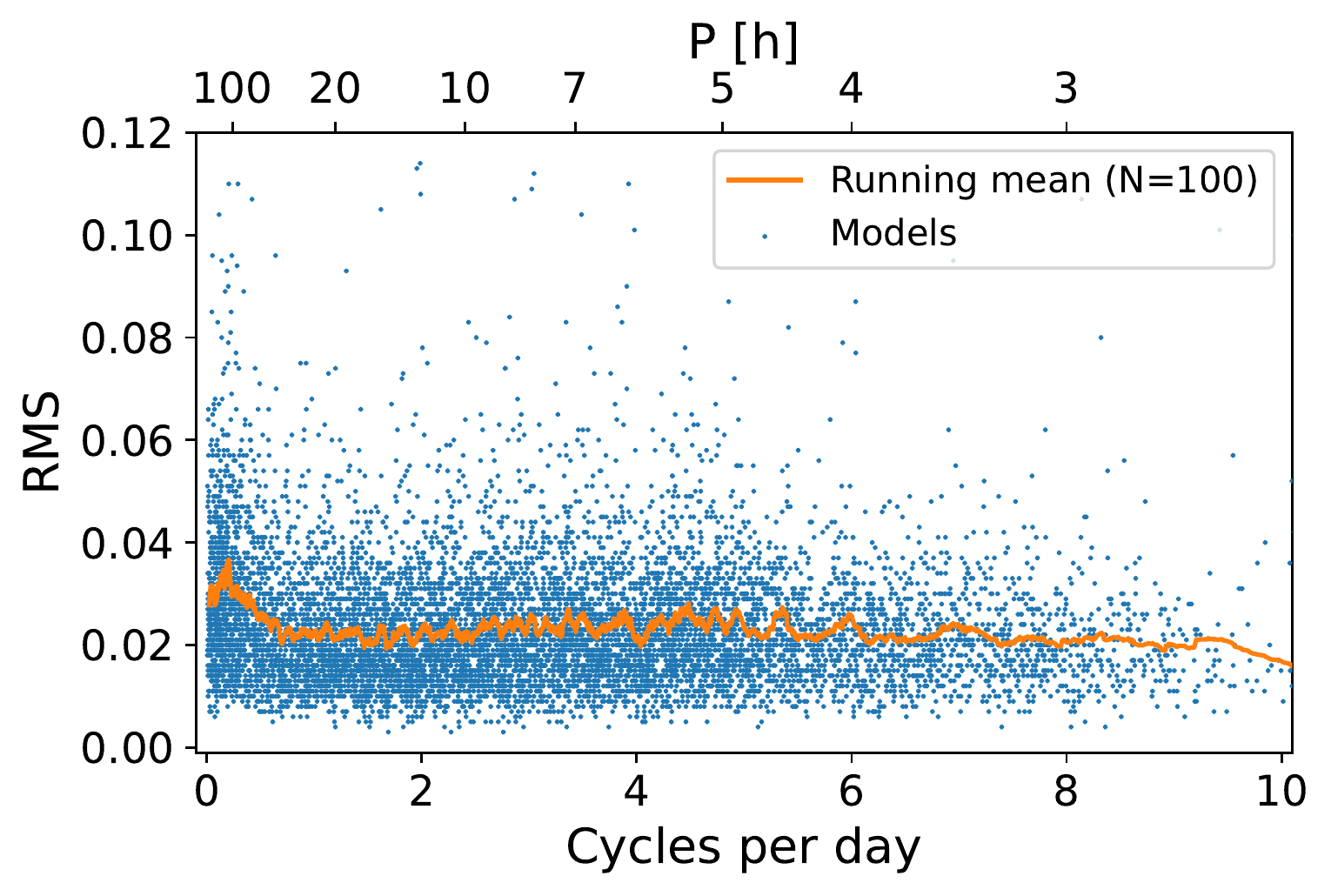}}\\
\end{center}
\caption{\label{fig:rms}RMS vs. rotation period for DR3 solutions. The orange line represents the running mean of 100 solutions.}
\end{figure*}

\begin{figure*}
\begin{center}
\resizebox{1.0\hsize}{!}{\includegraphics{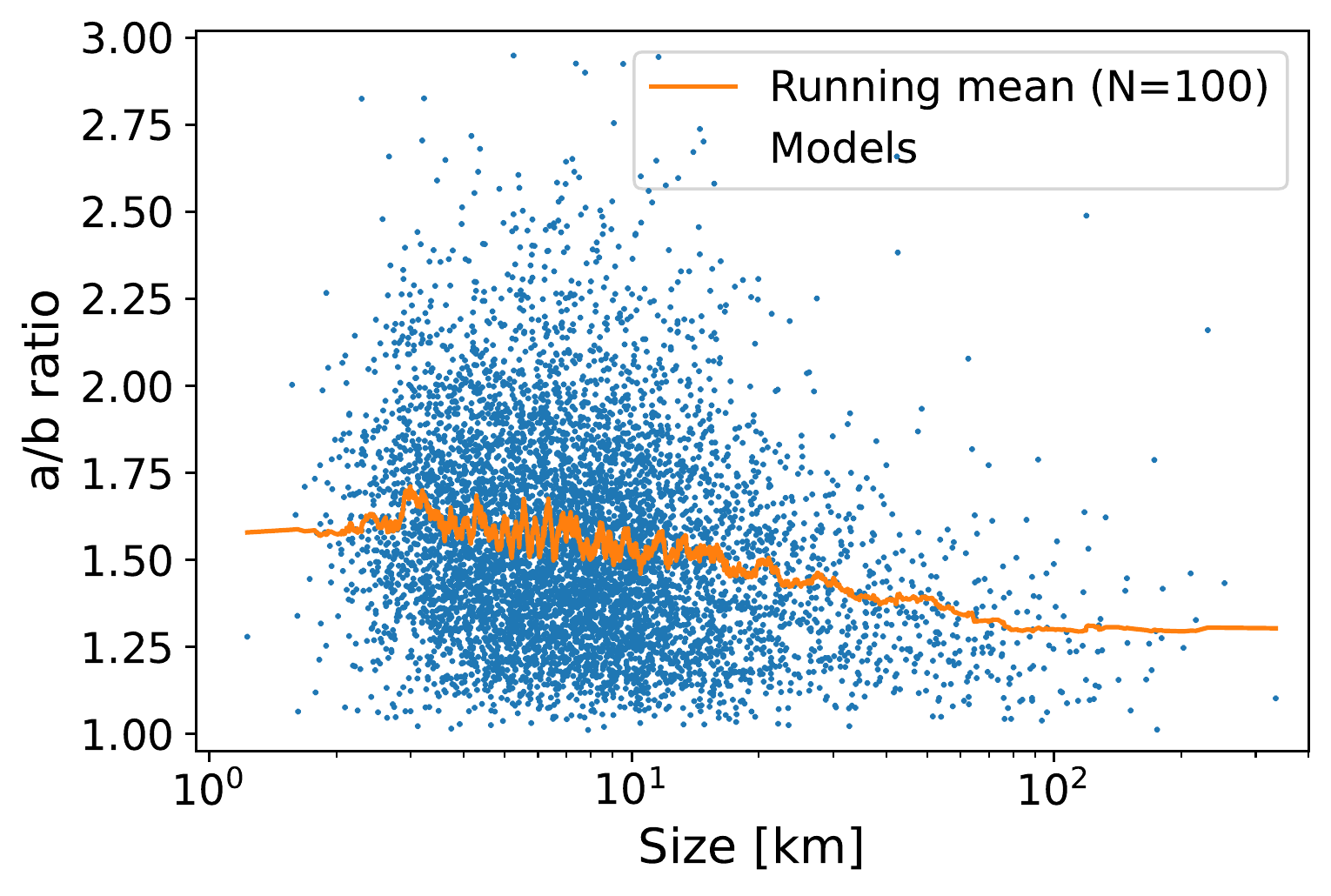}\includegraphics{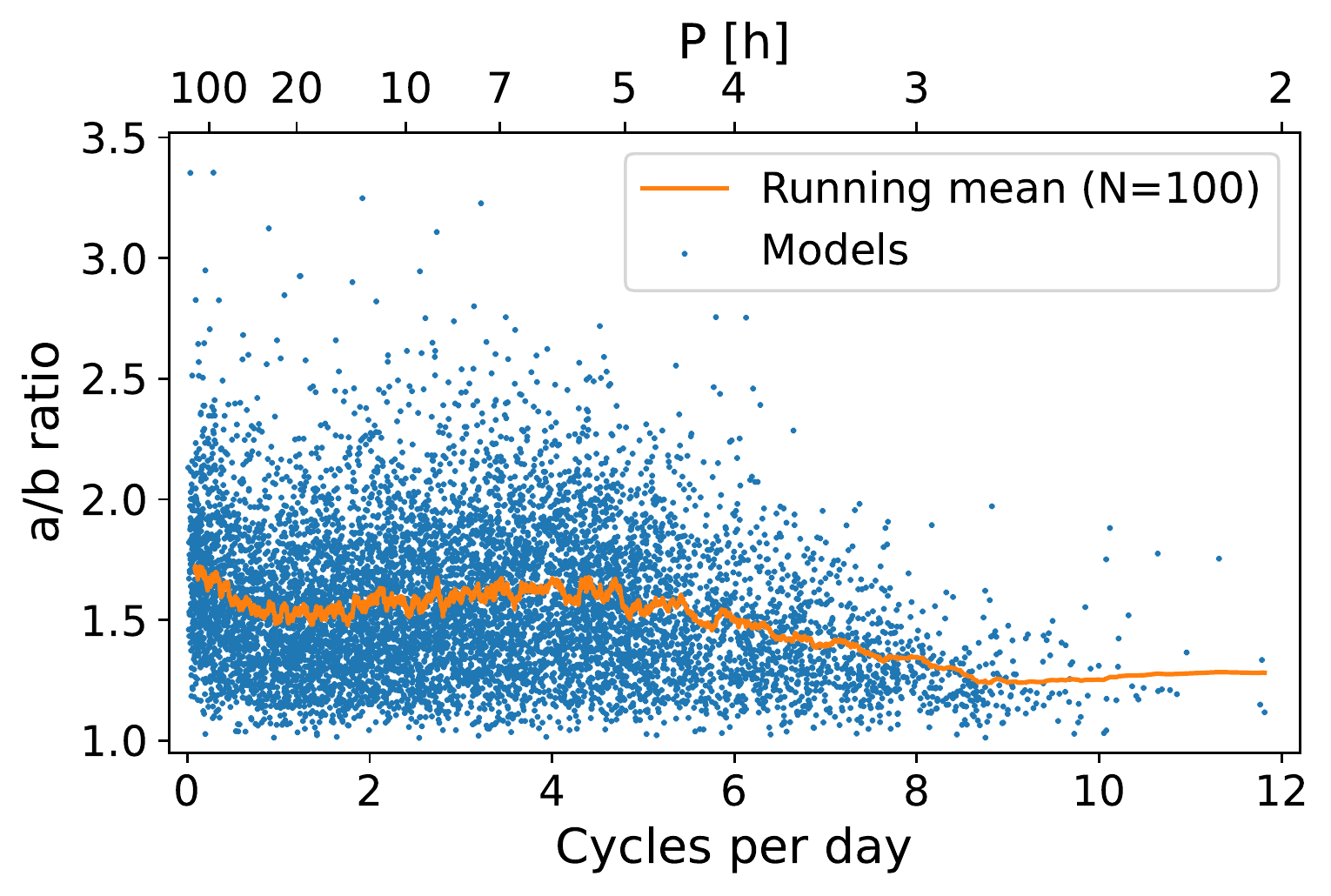}}\\
\end{center}
\caption{\label{fig:dab}Shape elongation $a/b$ as a function of asteroid size (volume-equivalent diameter) ({\it left\/}) and rotation period $P$ (in cycles per day and hours) ({\it right\/}). Larger asteroids are less elongated than smaller ones.  Faster rotators ($\gtrsim$8 cycles per day) are less elongated than the rest of the population. The orange line represents the running mean of 100 solutions.}
\end{figure*}

\begin{figure*}
\begin{center}
\resizebox{0.98\hsize}{!}{\includegraphics{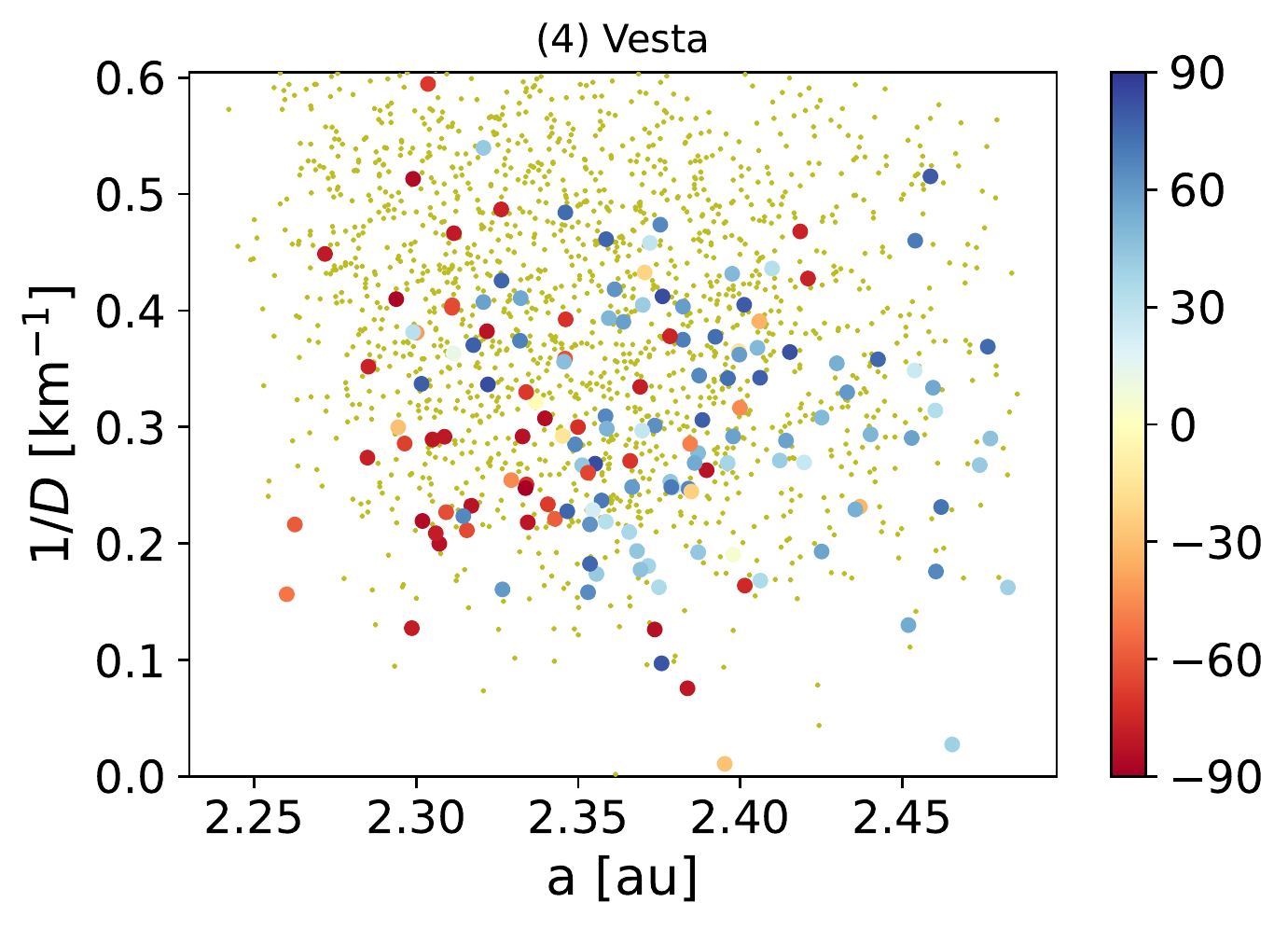}\includegraphics{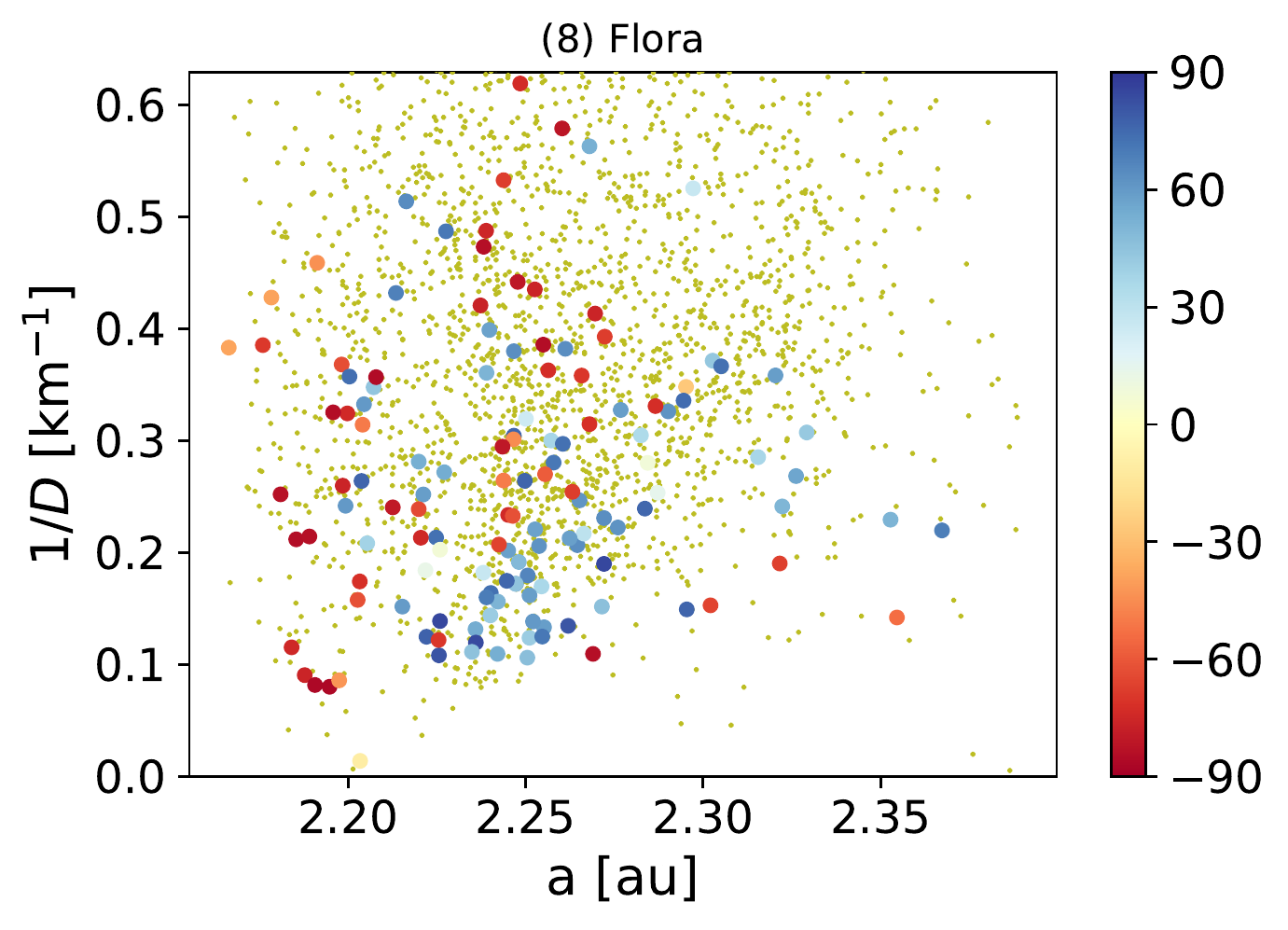}}\\
\resizebox{0.98\hsize}{!}{\includegraphics{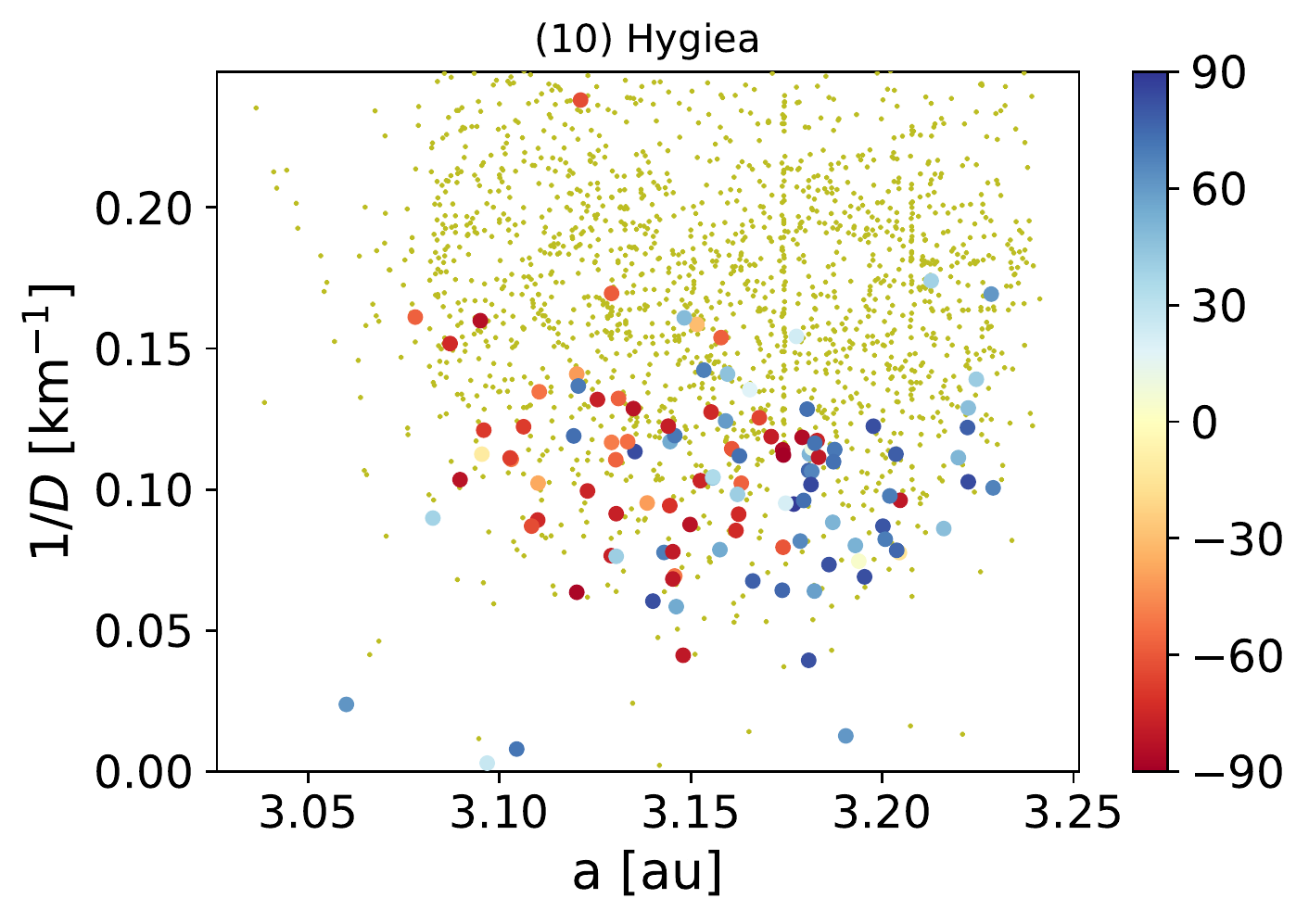}\includegraphics{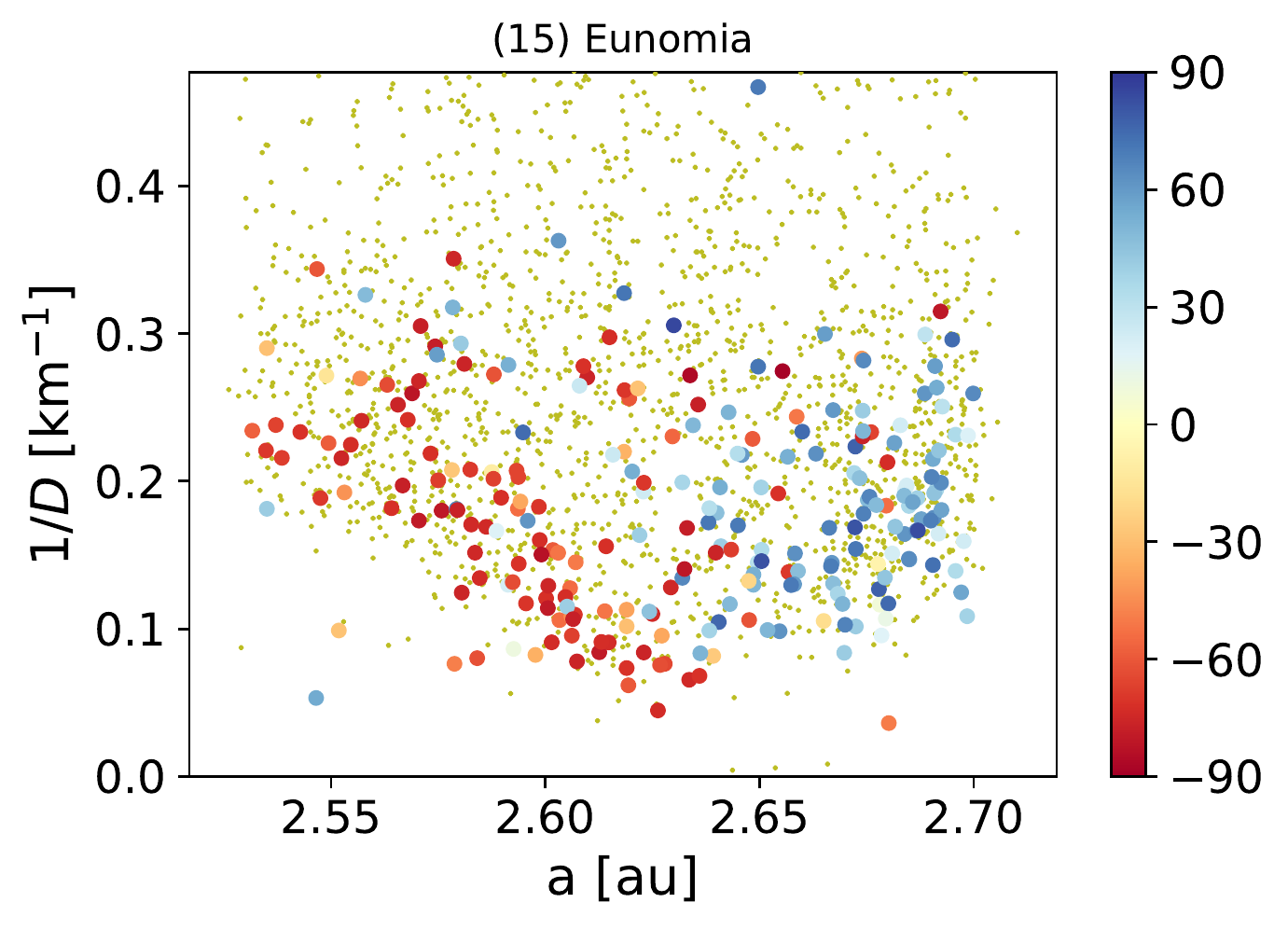}}\\
\resizebox{0.98\hsize}{!}{\includegraphics{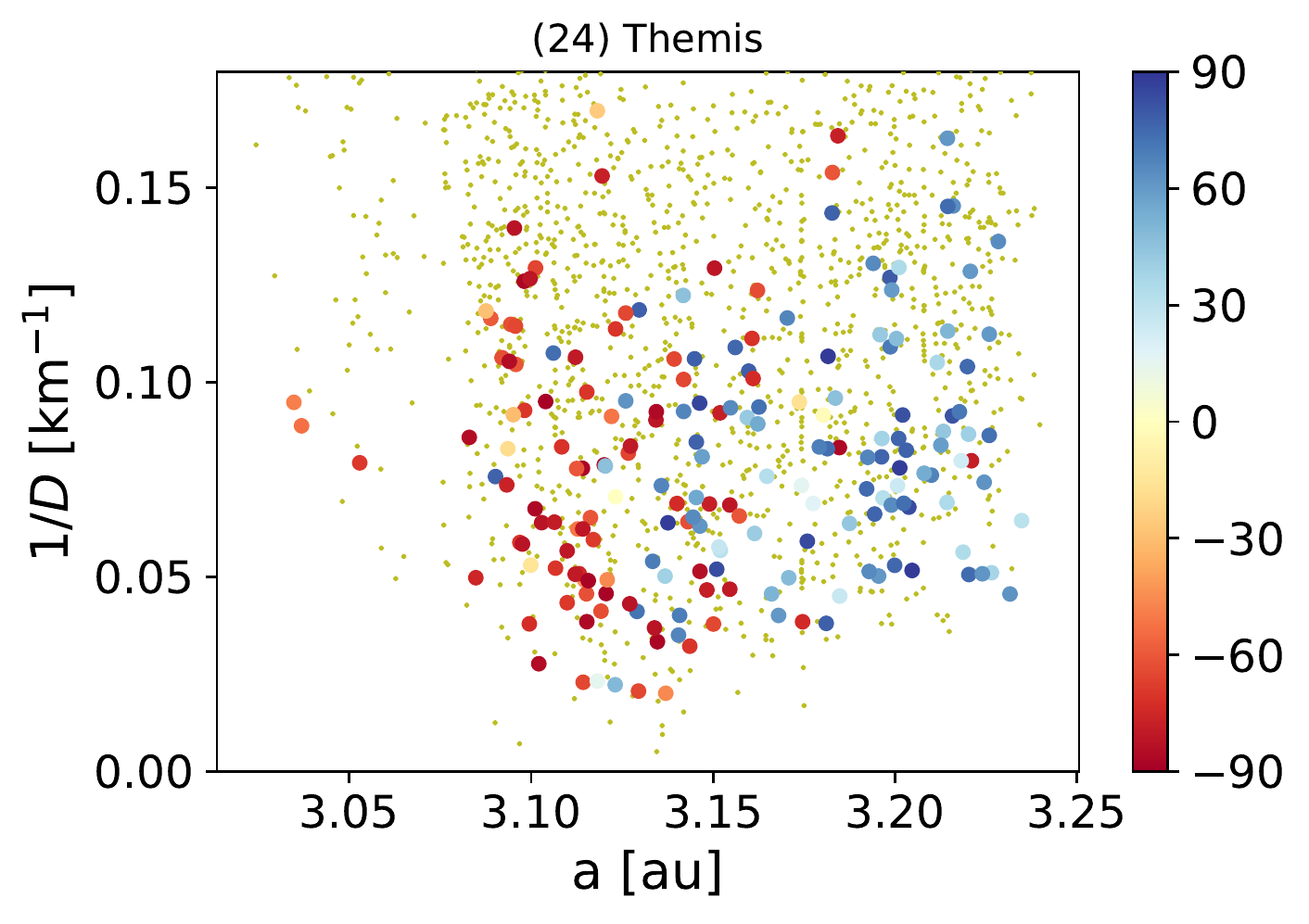}\includegraphics{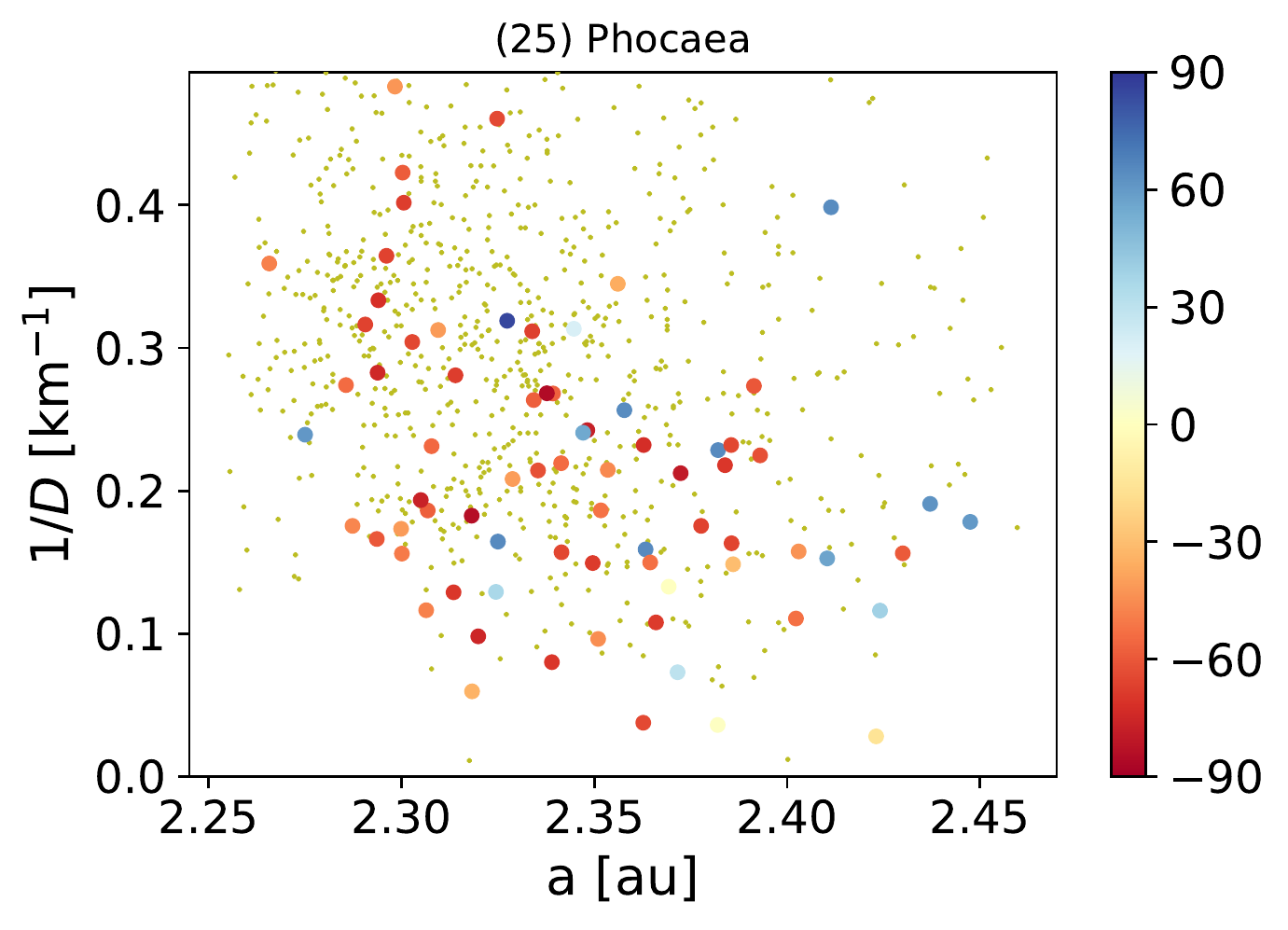}}\\
\end{center}
\caption{\label{fig:fam1} Color-coded spin axis distribution in asteroid families in proper semimajor axis and size. Family members for which we do not have a shape model are plotted as small dots.}
\end{figure*}

\begin{figure*}
\begin{center}
\resizebox{0.98\hsize}{!}{\includegraphics{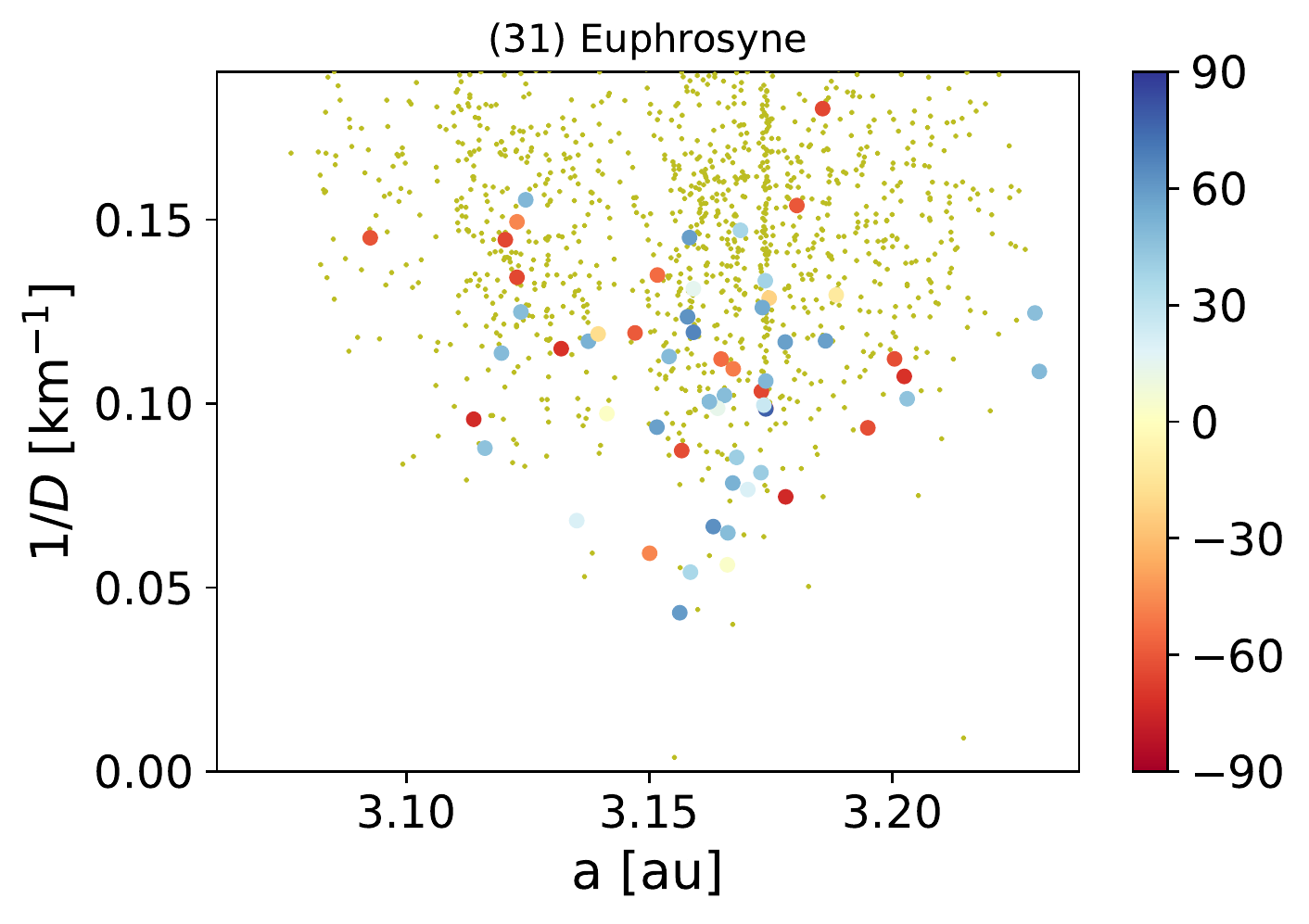}\includegraphics{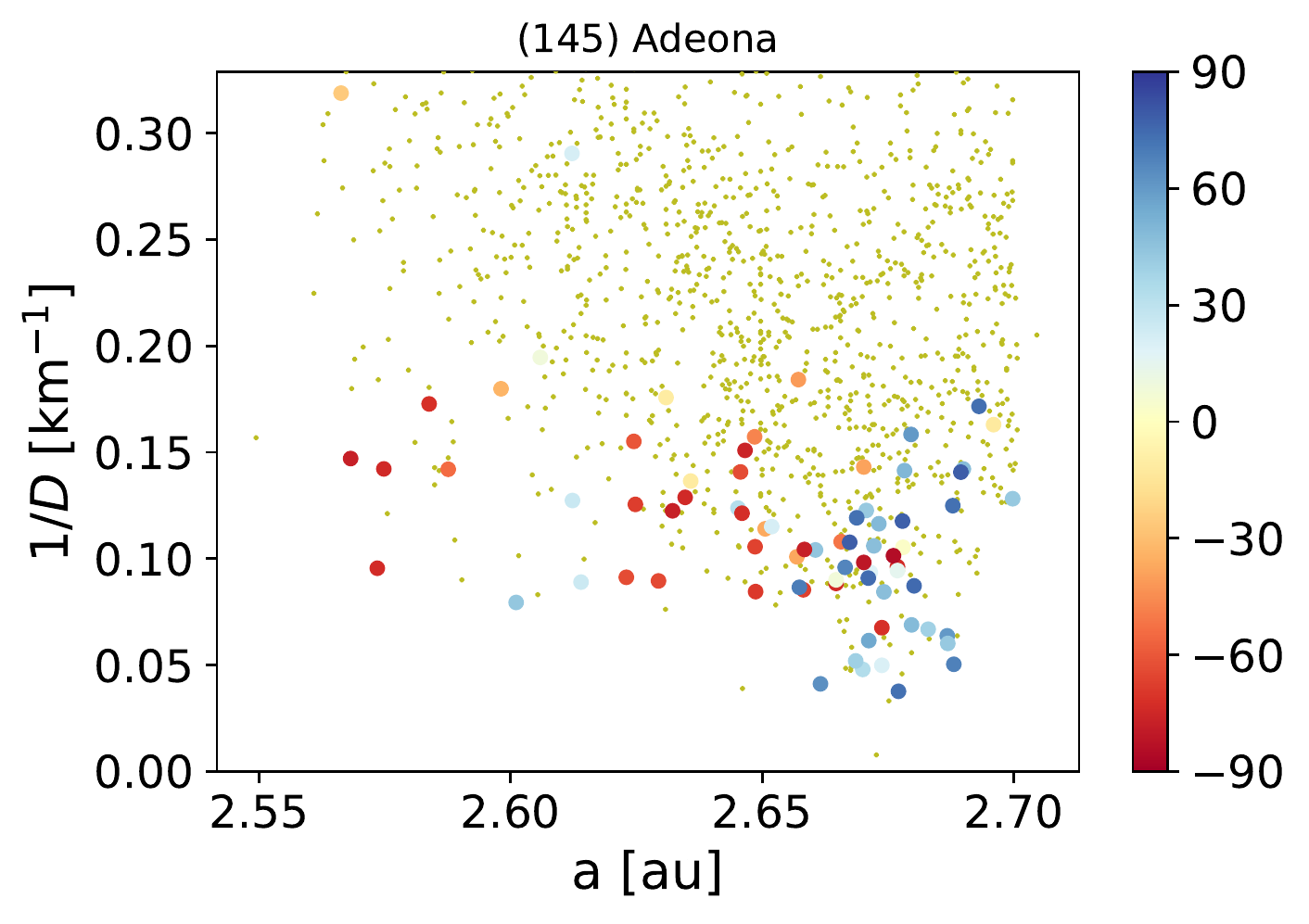}}\\
\resizebox{0.98\hsize}{!}{\includegraphics{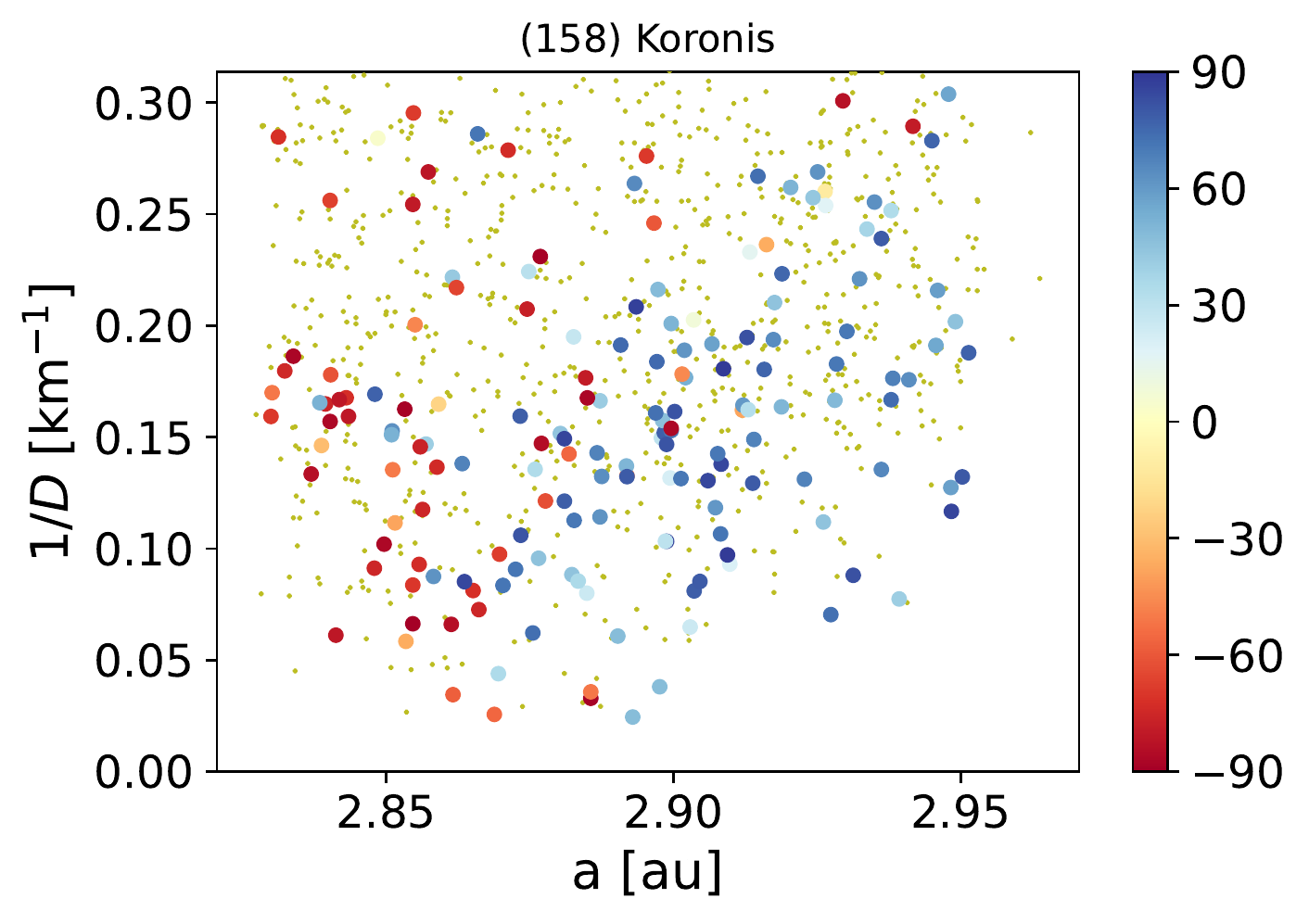}\includegraphics{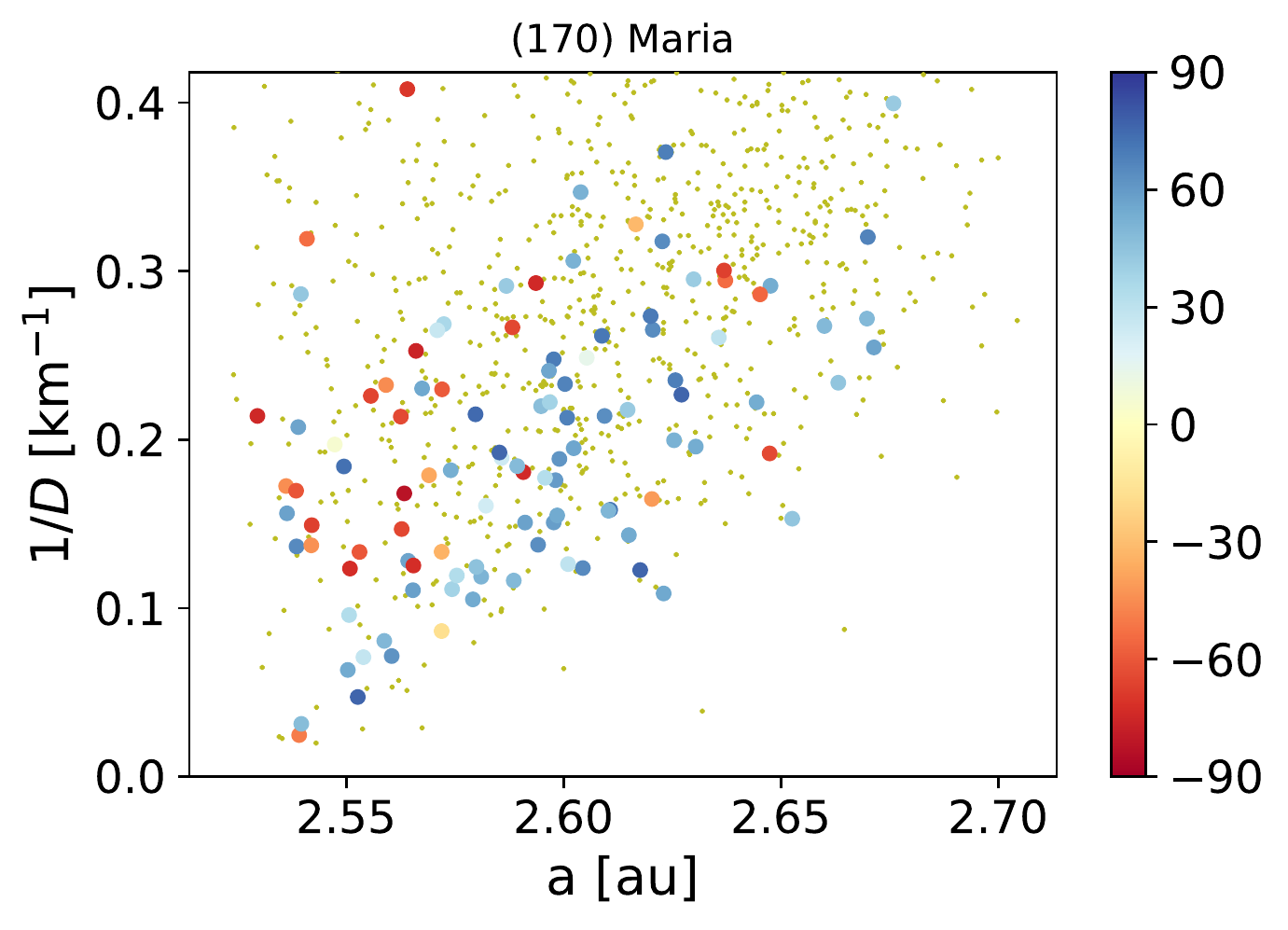}}\\
\resizebox{0.98\hsize}{!}{\includegraphics{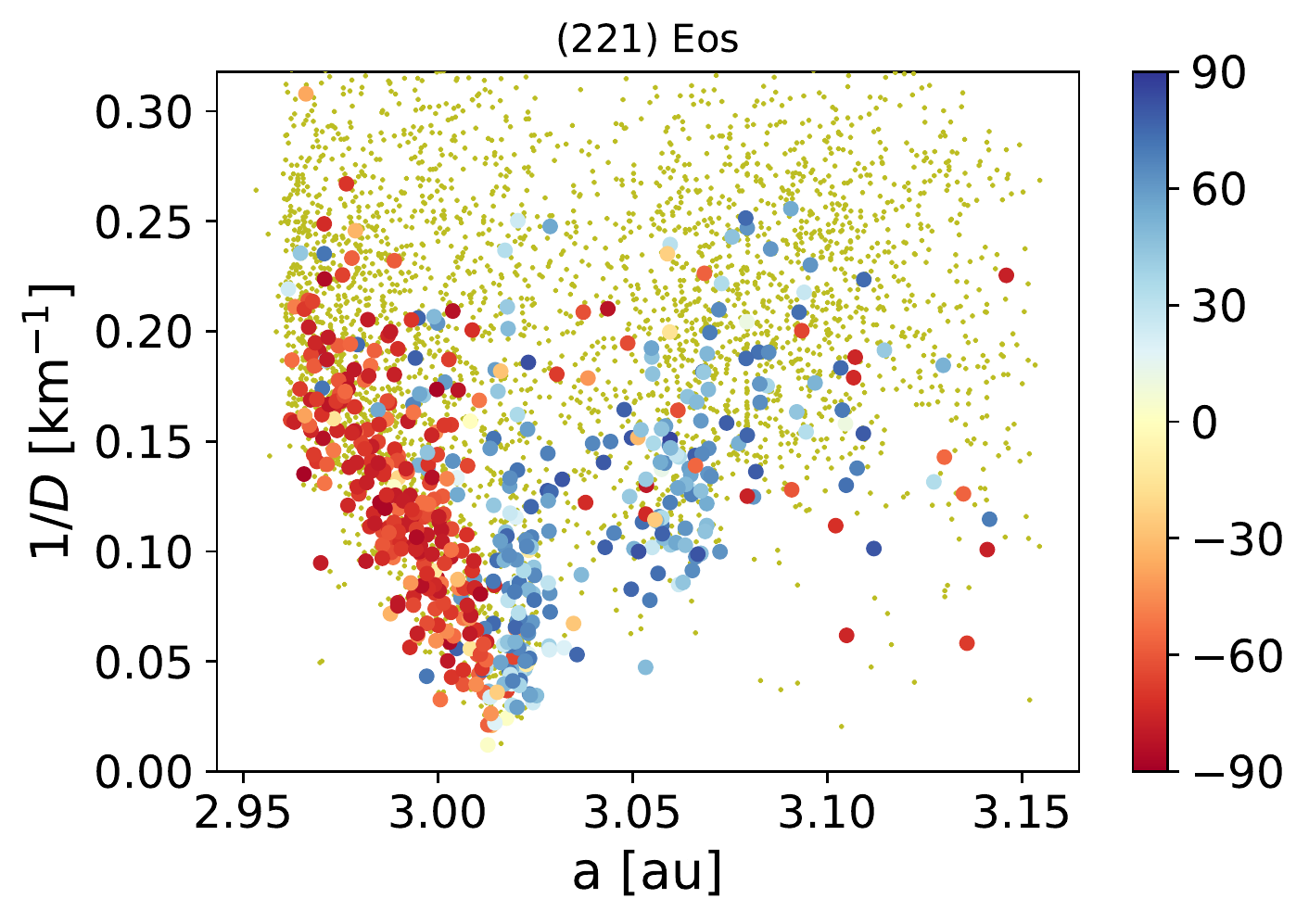}\includegraphics{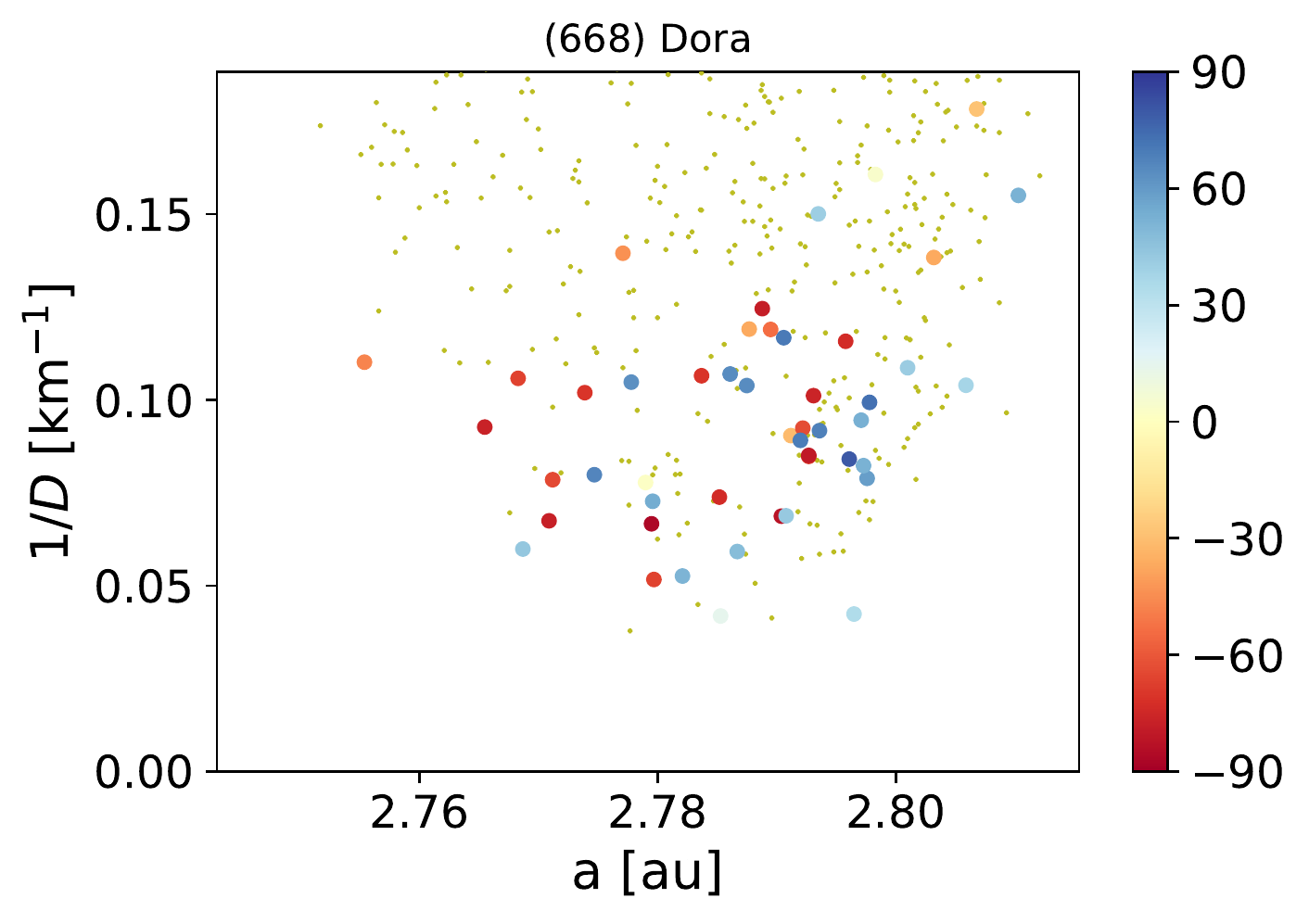}}\\
\end{center}
\caption{\label{fig:fam2} Color-coded spin axis distribution in asteroid families in proper semimajor axis and size. Family members for which we do not have a  shape model are plotted as small dots.}
\end{figure*}

\begin{figure*}
\begin{center}
\resizebox{0.98\hsize}{!}{\includegraphics{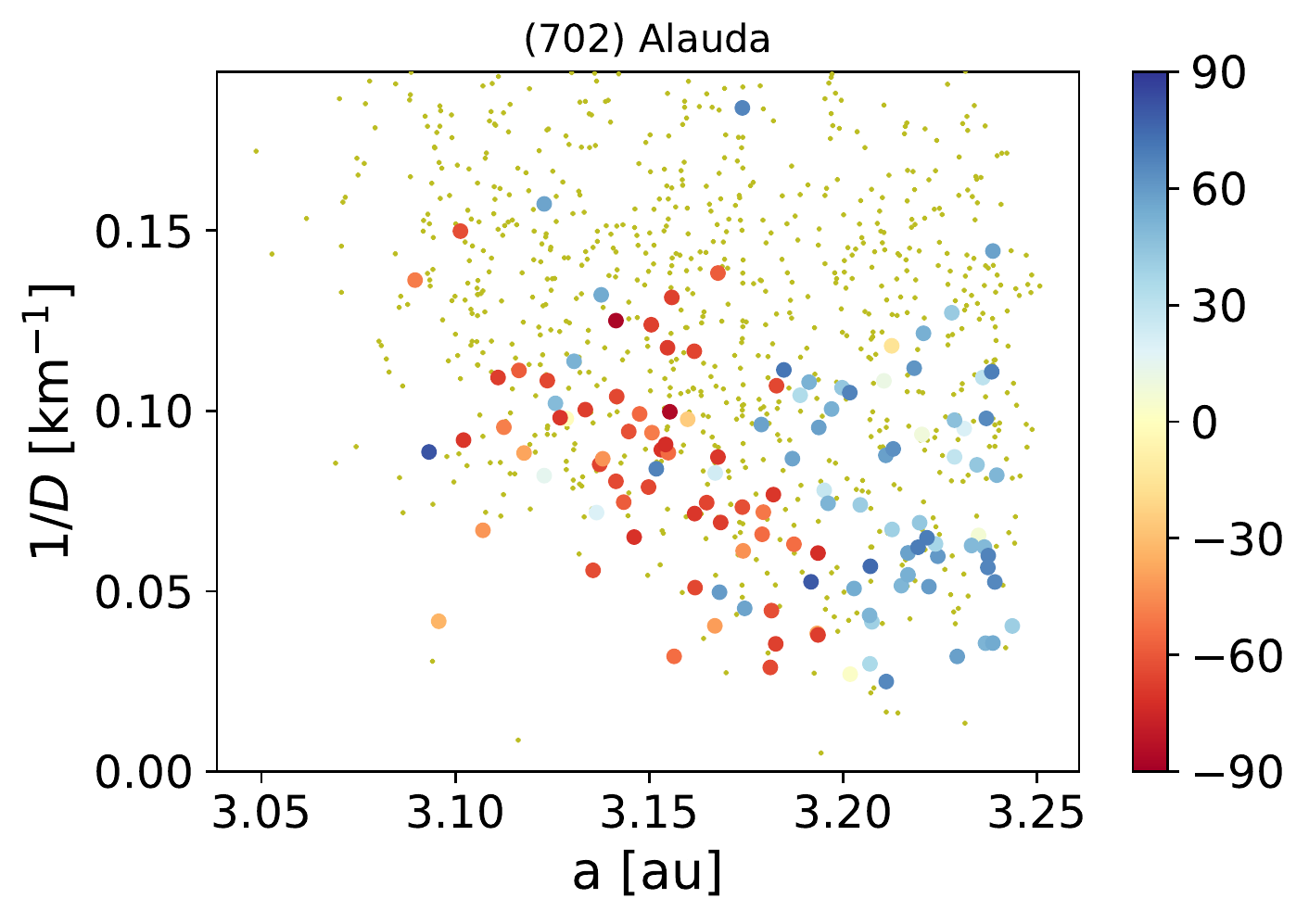}\includegraphics{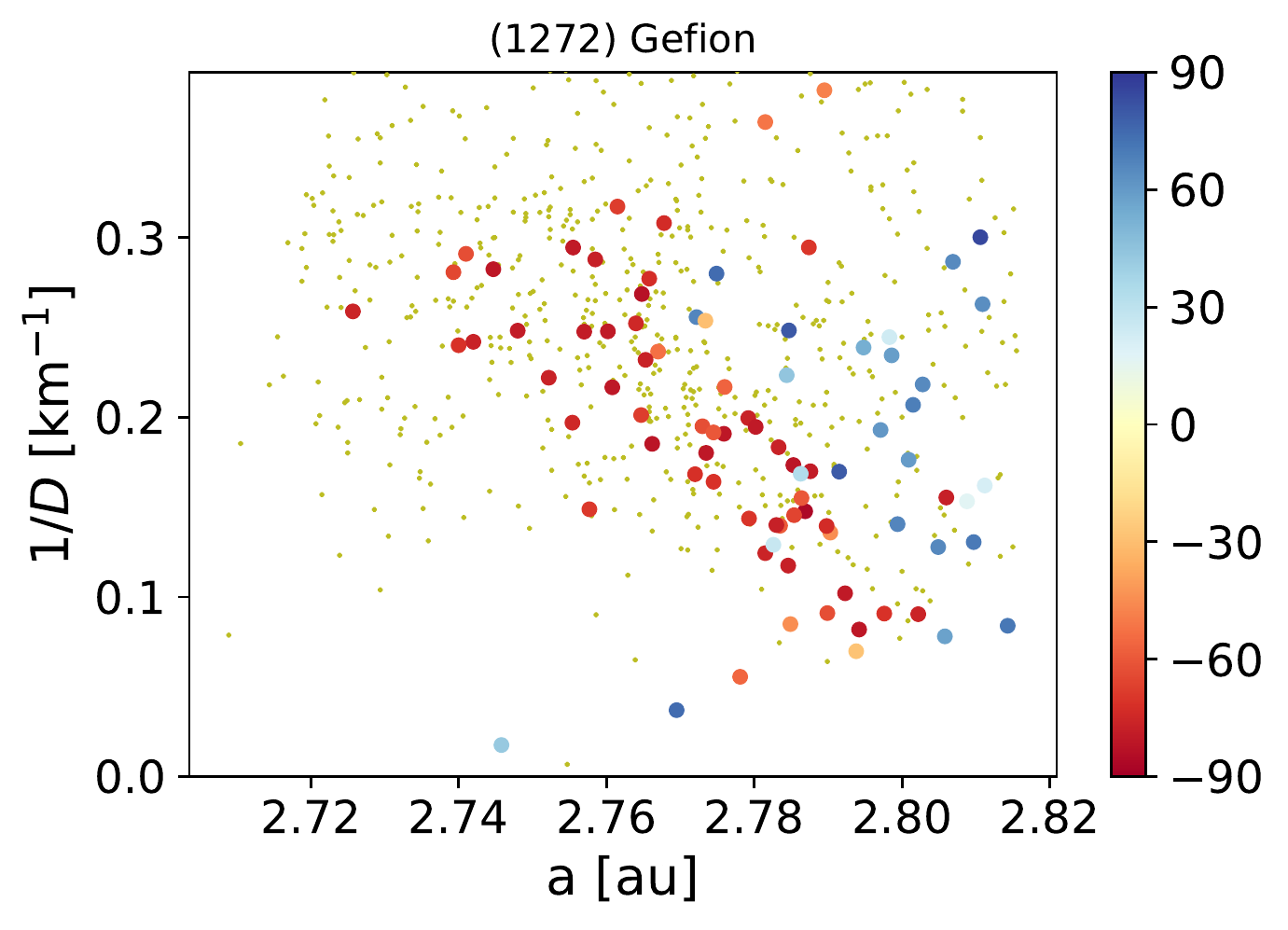}}\\
\end{center}
\caption{\label{fig:fam3} Color-coded spin axis distribution in asteroid families in proper semimajor axis and size. Family members for which we do not have a shape model are plotted as small dots.}
\end{figure*}

\begin{figure*}
\begin{center}
\resizebox{1.0\hsize}{!}{\includegraphics{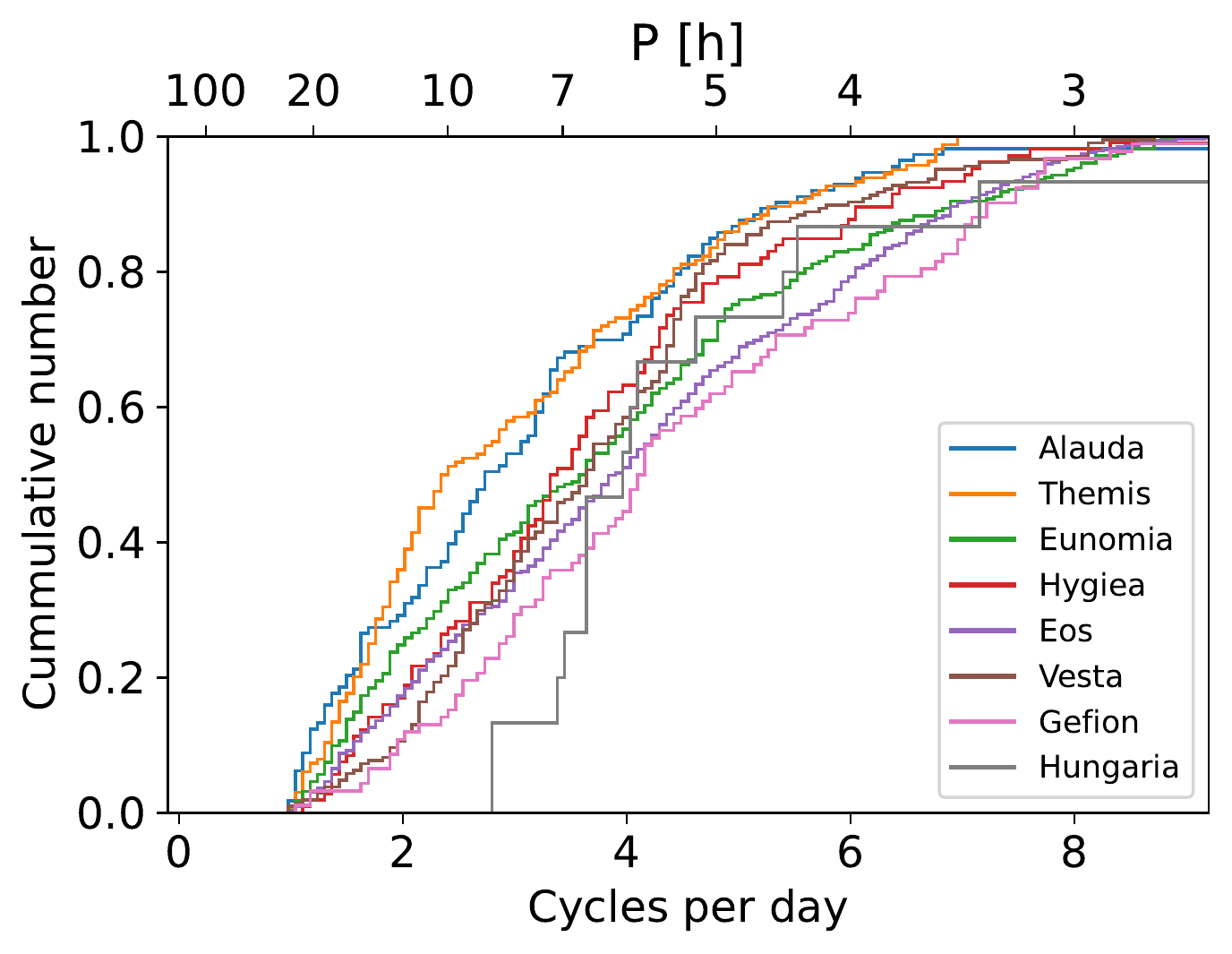}\includegraphics{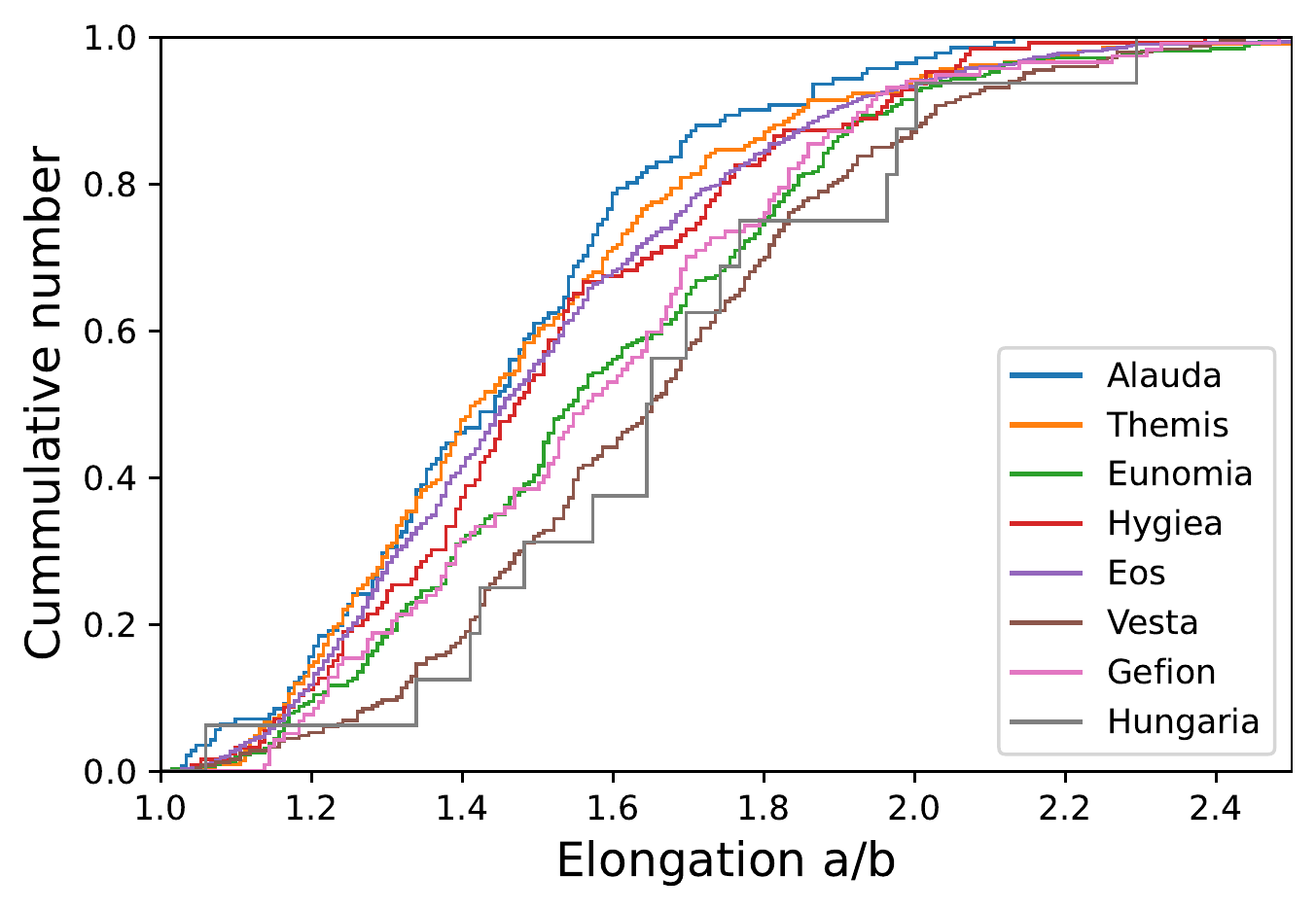}}\\
\end{center}
\caption{\label{fig:cumPerFam1} Cumulative distributions of periods ({\it left\/}) and elongations $a/b$ ({\it right\/}) in various families. They are ordered from approximately the oldest (Alauda, Themis) to the youngest (Gefion, Hungaria). Objects with periods $P>24$\,h were removed. Vesta, Gefion, and Hungaria have the lowest fraction of rotators with $P=10$--24\,h.}
\end{figure*}

\end{appendix}


\begin{thebibliography}{51}
\expandafter\ifx\csname natexlab\endcsname\relax\def\natexlab#1{#1}\fi

\bibitem[{{Babusiaux} {et~al.}(2022){Babusiaux}, {Fabricius}, {Khanna},
  {Muraveva}, {Reyl{\'e}}, {Spoto}, {Vallenari}, {Luri}, {Arenou}, {Alvarez},
  {Anders}, {Antoja}, {Balbinot}, {Barache}, {Bauchet}, {Bossini}, {Busonero},
  {Cantat-Gaudin}, {Carrasco}, {Dafonte}, {Diakite}, {Figueras},
  {Garcia-Gutierrez}, {Garofalo}, {Helmi}, {Jimenez-Arranz}, {Jordi},
  {Kervella}, {Kostrzewa-Rutkowska}, {Leclerc}, {Licata}, {Manteiga}, {Masip},
  {Monguio}, {Ramos}, {Robichon}, {Robin}, {Romero-Gomez}, {Saez}, {Santovena},
  {Spina}, {Torralba Elipe}, \& {Weiler}}]{Bab.ea:22}
{Babusiaux}, C., {Fabricius}, C., {Khanna}, S., {et~al.} 2022, arXiv e-prints,
  arXiv:2206.05989

\bibitem[{{Bembrick} {et~al.}(2002){Bembrick}, {Pereghy}, \&
  {Ainsworth}}]{Bem.ea:02}
{Bembrick}, C., {Pereghy}, B., \& {Ainsworth}, T. 2002, Minor Planet Bulletin,
  29, 21

\bibitem[{{Binzel}(1987)}]{Bin:87}
{Binzel}, R.~P. 1987, Icarus, 72, 135

\bibitem[{{Bottke} {et~al.}(2015){Bottke}, {Vokrouhlick{\'y}}, {Walsh},
  {Delbo}, {Michel}, {Lauretta}, {Campins}, {Connolly}, {Scheeres}, \&
  {Chelsey}}]{Bot.ea:15}
{Bottke}, W.~F., {Vokrouhlick{\'y}}, D., {Walsh}, K.~J., {et~al.} 2015,
  \icarus, 247, 191

\bibitem[{{Brinsfield}(2008)}]{Bri:08}
{Brinsfield}, J.~W. 2008, Minor Planet Bulletin, 35, 23

\bibitem[{{Carruba} {et~al.}(2017){Carruba}, {Vokrouhlick{\'y}}, \&
  {Nesvorn{\'y}}}]{Carruba2017}
{Carruba}, V., {Vokrouhlick{\'y}}, D., \& {Nesvorn{\'y}}, D. 2017, \mnras, 469,
  4400

\bibitem[{{Cibulkov{\'a}} {et~al.}(2016){Cibulkov{\'a}}, {{\v D}urech},
  {Vokrouhlick{\'y}}, {Kaasalainen}, \& {Oszkiewicz}}]{Cibulkova2016}
{Cibulkov{\'a}}, H., {{\v D}urech}, J., {Vokrouhlick{\'y}}, D., {Kaasalainen},
  M., \& {Oszkiewicz}, D.~A. 2016, \aap, 596, A57

\bibitem[{{Colazo} {et~al.}(2021){Colazo}, {Duffard}, \&
  {Weidmann}}]{Col.ea:21}
{Colazo}, M., {Duffard}, R., \& {Weidmann}, W. 2021, \mnras, 504, 761

\bibitem[{{Ditteon} {et~al.}(2018){Ditteon}, {Adam}, {Doyel}, {Gibson}, {Lee},
  {Linville}, {Michalik}, {Turner}, \& {Washburn}}]{Dit.ea:18}
{Ditteon}, R., {Adam}, A., {Doyel}, M., {et~al.} 2018, Minor Planet Bulletin,
  45, 13

\bibitem[{{Dose}(2021)}]{Dose:21}
{Dose}, E.~V. 2021, Minor Planet Bulletin, 48, 125

\bibitem[{{\v{D}urech} {et~al.}(2005){\v{D}urech}, {Grav}, {Jedicke},
  {Kaasalainen}, \& {Denneau}}]{Dur.ea:05}
{\v{D}urech}, J., {Grav}, T., {Jedicke}, R., {Kaasalainen}, M., \& {Denneau},
  L. 2005, Earth, Moon, and Planets, 97, 179

\bibitem[{{\v{D}urech} \& {Hanu{\v s}}(2018)}]{Dur.Han:18}
{\v{D}urech}, J. \& {Hanu{\v s}}, J. 2018, \aap, 620, A91

\bibitem[{{\v{D}urech} {et~al.}(2018){\v{D}urech}, {Hanu{\v s}}, \&
  {Al{\'{\i}}-Lagoa}}]{Dur.ea:18c}
{\v{D}urech}, J., {Hanu{\v s}}, J., \& {Al{\'{\i}}-Lagoa}, V. 2018, \aap, 617,
  A57

\bibitem[{{\v{D}urech} {et~al.}(2016){\v{D}urech}, {Hanu{\v s}}, {Oszkiewicz},
  \& {Van{\v c}o}}]{Dur.ea:16}
{\v{D}urech}, J., {Hanu{\v s}}, J., {Oszkiewicz}, D., \& {Van{\v c}o}, R. 2016,
  \aap, 587, A48

\bibitem[{{\v{D}urech} {et~al.}(2019){\v{D}urech}, {Hanu{\v{s}}}, \&
  {Van{\v{c}}o}}]{Dur.ea:19}
{\v{D}urech}, J., {Hanu{\v{s}}}, J., \& {Van{\v{c}}o}, R. 2019, \aap, 631, A2

\bibitem[{{\v{D}urech} {et~al.}(2010){\v{D}urech}, {Sidorin}, \&
  {Kaasalainen}}]{Dur.ea:10}
{\v{D}urech}, J., {Sidorin}, V., \& {Kaasalainen}, M. 2010, \aap, 513, A46

\bibitem[{{\v{D}urech} {et~al.}(2020){\v{D}urech}, {Tonry}, {Erasmus},
  {Denneau}, {Heinze}, {Flewelling}, \& {Van{\v{c}}o}}]{Dur.ea:20}
{\v{D}urech}, J., {Tonry}, J., {Erasmus}, N., {et~al.} 2020, \aap, 643, A59

\bibitem[{{Erasmus} {et~al.}(2020){Erasmus}, {Navarro-Meza}, {McNeill},
  {Trilling}, {Sickafoose}, {Denneau}, {Flewelling}, {Heinze}, \&
  {Tonry}}]{Era.ea:20}
{Erasmus}, N., {Navarro-Meza}, S., {McNeill}, A., {et~al.} 2020, \apjs, 247, 13

\bibitem[{{Fauvaud} \& {Fauvaud}(2013)}]{Fau.Fau:13}
{Fauvaud}, S. \& {Fauvaud}, M. 2013, Minor Planet Bulletin, 40, 224

\bibitem[{{Ferrero}(2021)}]{Fer:21}
{Ferrero}, A. 2021, Minor Planet Bulletin, 48, 7

\bibitem[{{Gaia Collaboration} {et~al.}(2018{\natexlab{a}}){Gaia
  Collaboration}, {Brown}, {Vallenari}, {Prusti}, {de Bruijne}, {Babusiaux},
  {Bailer-Jones}, {Biermann}, {Evans}, {Eyer}, \& et~al.}]{Gaia:18}
{Gaia Collaboration}, {Brown}, A.~G.~A., {Vallenari}, A., {et~al.}
  2018{\natexlab{a}}, \aap, 616, A1

\bibitem[{{Gaia Collaboration} {et~al.}(2016){Gaia Collaboration}, {Prusti},
  {de Bruijne}, {Brown}, {Vallenari}, {Babusiaux}, {Bailer-Jones}, {Bastian},
  {Biermann}, {Evans}, \& et~al.}]{Gaia:16}
{Gaia Collaboration}, {Prusti}, T., {de Bruijne}, J.~H.~J., {et~al.} 2016,
  \aap, 595, A1

\bibitem[{{Gaia Collaboration} {et~al.}(2018{\natexlab{b}}){Gaia
  Collaboration}, {Spoto}, {Tanga}, {Mignard}, {Berthier}, {Carry}, {Cellino},
  {Dell'Oro}, {Hestroffer}, {Muinonen}, \& et~al.}]{Spo.ea:18}
{Gaia Collaboration}, {Spoto}, F., {Tanga}, P., {et~al.} 2018{\natexlab{b}},
  \aap, 616, A13

\bibitem[{{Hanu{\v s}} {et~al.}(2018){Hanu{\v s}}, {Delbo'},
  {Al{\'{\i}}-Lagoa}, {Bolin}, {Jedicke}, {{\v D}urech}, {Cibulkov{\'a}},
  {Pravec}, {Ku{\v s}nir{\'a}k}, {Behrend}, {Marchis}, {Antonini}, {Arnold},
  {Audejean}, {Bachschmidt}, {Bernasconi}, {Brunetto}, {Casulli}, {Dymock},
  {Esseiva}, {Esteban}, {Gerteis}, {de Groot}, {Gully}, {Hamanowa}, {Hamanowa},
  {Krafft}, {Lehk{\'y}}, {Manzini}, {Michelet}, {Morelle}, {Oey}, {Pilcher},
  {Reignier}, {Roy}, {Salom}, \& {Warner}}]{Han.ea:18a}
{Hanu{\v s}}, J., {Delbo'}, M., {Al{\'{\i}}-Lagoa}, V., {et~al.} 2018, \icarus,
  299, 84

\bibitem[{{Hanu{\v s}} {et~al.}(2013){Hanu{\v s}}, {{\v D}urech}, {Bro{\v z}},
  {Marciniak}, {Warner}, {Pilcher}, {Stephens}, {Behrend}, {Carry}, {{\v
  C}apek}, {Antonini}, {Audejean}, {Augustesen}, {Barbotin}, {Baudouin},
  {Bayol}, {Bernasconi}, {Borczyk}, {Bosch}, {Brochard}, {Brunetto}, {Casulli},
  {Cazenave}, {Charbonnel}, {Christophe}, {Colas}, {Coloma}, {Conjat},
  {Cooney}, {Correira}, {Cotrez}, {Coupier}, {Crippa}, {Cristofanelli},
  {Dalmas}, {Danavaro}, {Demeautis}, {Droege}, {Durkee}, {Esseiva}, {Esteban},
  {Fagas}, {Farroni}, {Fauvaud}, {Fauvaud}, {Del Freo}, {Garcia}, {Geier},
  {Godon}, {Grangeon}, {Hamanowa}, {Hamanowa}, {Heck}, {Hellmich}, {Higgins},
  {Hirsch}, {Husarik}, {Itkonen}, {Jade}, {Kami{\'n}ski}, {Kankiewicz},
  {Klotz}, {Koff}, {Kryszczy{\'n}ska}, {Kwiatkowski}, {Laffont}, {Leroy},
  {Lecacheux}, {Leonie}, {Leyrat}, {Manzini}, {Martin}, {Masi}, {Matter},
  {Micha{\l}owski}, {Micha{\l}owski}, {Micha{\l}owski}, {Michelet},
  {Michelsen}, {Morelle}, {Mottola}, {Naves}, {Nomen}, {Oey}, {Og{\l}oza},
  {Oksanen}, {Oszkiewicz}, {P{\"a}{\"a}kk{\"o}nen}, {Paiella}, {Pallares},
  {Paulo}, {Pavic}, {Payet}, {Poli{\'n}ska}, {Polishook}, {Poncy}, {Revaz},
  {Rinner}, {Rocca}, {Roche}, {Romeuf}, {Roy}, {Saguin}, {Salom}, {Sanchez},
  {Santacana}, {Santana-Ros}, {Sareyan}, {Sobkowiak}, {Sposetti}, {Starkey},
  {Stoss}, {Strajnic}, {Teng}, {Tr{\'e}gon}, {Vagnozzi}, {Velichko},
  {Waelchli}, {Wagrez}, \& {W{\"u}cher}}]{Han.ea:13b}
{Hanu{\v s}}, J., {{\v D}urech}, J., {Bro{\v z}}, M., {et~al.} 2013, \aap, 551,
  A67

\bibitem[{{Hanu\v{s}} {et~al.}(2011){Hanu\v{s}}, {\v{D}urech}, {Bro\v{z}},
  {Warner}, {Pilcher}, {Stephens}, {Oey}, {Bernasconi}, {Casulli}, {Behrend},
  {Polishook}, {Henych}, {Lehk{\'y}}, {Yoshida}, \& {Ito}}]{Hanus2011}
{Hanu\v{s}}, J., {\v{D}urech}, J., {Bro\v{z}}, M., {et~al.} 2011, \aap, 530,
  A134

\bibitem[{{Kaasalainen}(2004)}]{Kaa:04}
{Kaasalainen}, M. 2004, \aap, 422, L39

\bibitem[{{Kaasalainen} \& {Lamberg}(2006)}]{Kaa.Lam:06}
{Kaasalainen}, M. \& {Lamberg}, L. 2006, Inverse Problems, 22, 749

\bibitem[{{Kaasalainen} {et~al.}(2002){Kaasalainen}, {Mottola}, \&
  {Fulchignomi}}]{Kaa.ea:02c}
{Kaasalainen}, M., {Mottola}, S., \& {Fulchignomi}, M. 2002, in
  {Asteroids~III}, ed. W.~F. {Bottke}, A.~{Cellino}, P.~{Paolicchi}, \& R.~P.
  {Binzel} (Tucson: {University of Arizona Press}), 139--150

\bibitem[{{Kaasalainen} \& {Torppa}(2001)}]{Kaa.Tor:01}
{Kaasalainen}, M. \& {Torppa}, J. 2001, Icarus, 153, 24

\bibitem[{{Kaasalainen} {et~al.}(2001){Kaasalainen}, {Torppa}, \&
  {Muinonen}}]{Kaasalainen2001b}
{Kaasalainen}, M., {Torppa}, J., \& {Muinonen}, K. 2001, Icarus, 153, 37

\bibitem[{{Kalup} {et~al.}(2021){Kalup}, {Moln{\'a}r}, {Kiss}, {Szab{\'o}},
  {P{\'a}l}, {Szak{\'a}ts}, {S{\'a}rneczky}, {Vink{\'o}}, {Szab{\'o}},
  {Kecskem{\'e}thy}, \& {Kiss}}]{Kalup2021}
{Kalup}, C.~E., {Moln{\'a}r}, L., {Kiss}, C., {et~al.} 2021, \apjs, 254, 7

\bibitem[{{Lamberg} \& {Kaasalainen}(2001)}]{Lam.Kaa:01}
{Lamberg}, L. \& {Kaasalainen}, M. 2001, J. Comp. Appl. Math., 137, 213

\bibitem[{{Lecrone} {et~al.}(2004){Lecrone}, {Duncan}, \&
  {Ditteon}}]{Lec.ea:04}
{Lecrone}, C., {Duncan}, A., \& {Ditteon}, R. 2004, Minor Planet Bulletin, 31,
  78

\bibitem[{{Marciniak} {et~al.}(2015){Marciniak}, {Pilcher}, {Oszkiewicz},
  {Santana-Ros}, {Urakawa}, {Fauvaud}, {Kankiewicz}, {Tychoniec}, {Fauvaud},
  {Hirsch}, {Horbowicz}, {Kami{\'n}ski}, {Konstanciak}, {Kosturkiewicz},
  {Murawiecka}, {Nadolny}, {Nishiyama}, {Okumura}, {Poli{\'n}ska}, {Richard},
  {Sakamoto}, {Sobkowiak}, {Stachowski}, \& {Trela}}]{Mar.ea:15}
{Marciniak}, A., {Pilcher}, F., {Oszkiewicz}, D., {et~al.} 2015, \planss, 118,
  256

\bibitem[{{Marciniak} {et~al.}(2021){Marciniak}, {{\v{D}}urech},
  {Al{\'\i}-Lagoa}, {Og{\l}oza}, {Szak{\'a}ts}, {M{\"u}ller}, {Moln{\'a}r},
  {P{\'a}l}, {Monteiro}, {Arcoverde}, {Behrend}, {Benkhaldoun}, {Bernasconi},
  {Bosch}, {Brincat}, {Brunetto}, {Butkiewicz-B{\k{a}}k}, {Del Freo},
  {Duffard}, {Evangelista-Santana}, {Farroni}, {Fauvaud}, {Fauvaud}, {Ferrais},
  {Geier}, {Golonka}, {Grice}, {Hirsch}, {Horbowicz}, {Jehin}, {Julien},
  {Kalup}, {Kami{\'n}ski}, {Kami{\'n}ska}, {Kankiewicz}, {Kecskem{\'e}thy},
  {Kim}, {Kim}, {Konstanciak}, {Krajewski}, {Kudak}, {Kulczak}, {Kundera},
  {Lazzaro}, {Manzini}, {Medeiros}, {Michimani-Garcia}, {Morales}, {Nadolny},
  {Oszkiewicz}, {Pak{\v{s}}tien{\.{e}}}, {Paw{\l}owski}, {Perig}, {Pilcher},
  {Pinel}, {Podlewska-Gaca}, {Polakis}, {Richard}, {Rodrigues}, {Rond{\'o}n},
  {Roy}, {Sanabria}, {Santana-Ros}, {Skiff}, {Skrzypek}, {Sobkowiak}, {Sonbas},
  {Stachowski}, {Strajnic}, {Trela}, {Tychoniec}, {Urakawa}, {Verebelyi},
  {Wagrez}, {{\.Z}ejmo}, \& {{\.Z}ukowski}}]{Mar.ea:21}
{Marciniak}, A., {{\v{D}}urech}, J., {Al{\'\i}-Lagoa}, V., {et~al.} 2021, \aap,
  654, A87

\bibitem[{{Mommert} {et~al.}(2018){Mommert}, {McNeill}, {Trilling},
  {Moskovitz}, \& {Delbo'}}]{Mom.ea:18}
{Mommert}, M., {McNeill}, A., {Trilling}, D.~E., {Moskovitz}, N., \& {Delbo'},
  M. 2018, \aj, 156, 139

\bibitem[{{Nesvorn{\'y}} {et~al.}(2015){Nesvorn{\'y}}, {Bro{\v z}}, \&
  {Carruba}}]{Nesvorny2015}
{Nesvorn{\'y}}, D., {Bro{\v z}}, M., \& {Carruba}, V. 2015, {Identification and
  Dynamical Properties of Asteroid Families}, ed. P.~{Michel}, F.~E. {DeMeo},
  \& W.~F. {Bottke}, 297--321

\bibitem[{{Ostro} \& {Connelly}(1984)}]{Ost.Con:84}
{Ostro}, S.~J. \& {Connelly}, R. 1984, Icarus, 57, 443

\bibitem[{{P{\'a}l} {et~al.}(2020){P{\'a}l}, {Szak{\'a}ts}, {Kiss}, {B{\'o}di},
  {Bogn{\'a}r}, {Kalup}, {Kiss}, {Marton}, {Moln{\'a}r}, {Plachy},
  {S{\'a}rneczky}, {Szab{\'o}}, \& {Szab{\'o}}}]{Pal.ea:20}
{P{\'a}l}, A., {Szak{\'a}ts}, R., {Kiss}, C., {et~al.} 2020, \apjs, 247, 26

\bibitem[{{Ryan} {et~al.}(2017){Ryan}, {Sharkey}, \& {Woodward}}]{Rya.ea:17}
{Ryan}, E.~L., {Sharkey}, B. N.~L., \& {Woodward}, C.~E. 2017, \aj, 153, 116

\bibitem[{{Statler}(2009)}]{Sta:09}
{Statler}, T.~S. 2009, \icarus, 202, 502

\bibitem[{{Szab{\'o}} {et~al.}(2017){Szab{\'o}}, {P{\'a}l}, {Kiss}, {Kiss},
  {Moln{\'a}r}, {Hanyecz}, {Plachy}, {S{\'a}rneczky}, \&
  {Szab{\'o}}}]{Sza.ea:17}
{Szab{\'o}}, G.~M., {P{\'a}l}, A., {Kiss}, C., {et~al.} 2017, \aap, 599, A44

\bibitem[{{Szab{\'o}} {et~al.}(2022){Szab{\'o}}, {P{\'a}l}, {Szigeti},
  {Bogn{\'a}r}, {B{\'o}di}, {Kalup}, {J{\"a}ger}, {Kiss}, {Kiss}, {Kov{\'a}cs},
  {Marton}, {Moln{\'a}r}, {Plachy}, {S{\'a}rneczky}, {Szak{\'a}ts}, \&
  {Szab{\'o}}}]{Szabo2022}
{Szab{\'o}}, G.~M., {P{\'a}l}, A., {Szigeti}, L., {et~al.} 2022, \aap, 661, A48

\bibitem[{{Tanga} {et~al.}(2022){Tanga}, {Pauwels}, {Mignard}, {Muinonen},
  {Cellino}, {David}, {Hestroffer}, {Spoto}, {Berthier}, {Guiraud}, {Roux},
  {Carry}, {Delbo}, {Dell Oro}, {Fouron}, {Galluccio}, {Jonckheere}, {Klioner},
  {Lefustec}, {Liberato}, {Ord{\'e}novic}, {Oreshina-Slezak}, {Penttil{\"a}},
  {Pailler}, {Panem}, {Petit}, {Portell}, {Poujoulet}, {Thuillot}, {Van
  Hemelryck}, {Burlacu}, {Lasne}, \& {Managa}}]{Tan.ea:22}
{Tanga}, P., {Pauwels}, T., {Mignard}, F., {et~al.} 2022, arXiv e-prints,
  arXiv:2206.05561

\bibitem[{{Tsiganis} {et~al.}(2007){Tsiganis}, {Kne{\v{z}}evi{\'c}}, \&
  {Varvoglis}}]{Tsiganis2007}
{Tsiganis}, K., {Kne{\v{z}}evi{\'c}}, Z., \& {Varvoglis}, H. 2007, \icarus,
  186, 484

\bibitem[{{Warner} {et~al.}(2009){Warner}, {Harris}, \& {Pravec}}]{War.ea:09}
{Warner}, B.~D., {Harris}, A.~W., \& {Pravec}, P. 2009, Icarus, 202, 134

\bibitem[{{Warner} \& {Stephens}(2021)}]{War.Ste:21b}
{Warner}, B.~D. \& {Stephens}, R.~D. 2021, Minor Planet Bulletin, 48, 17

\bibitem[{{Waszczak} {et~al.}(2015){Waszczak}, {Chang}, {Ofek}, {Laher},
  {Masci}, {Levitan}, {Surace}, {Cheng}, {Ip}, {Kinoshita}, {Helou}, {Prince},
  \& {Kulkarni}}]{Was.ea:15}
{Waszczak}, A., {Chang}, C.-K., {Ofek}, E.~O., {et~al.} 2015, \aj, 150, 75

\bibitem[{{Wilawer} {et~al.}(2022){Wilawer}, {Oszkiewicz}, {Kryszczy{\'n}ska},
  {Marciniak}, {Shevchenko}, {Belskaya}, {Kwiatkowski}, {Kankiewicz},
  {Horbowicz}, {Kudak}, {Kulczak}, {Perig}, \& {Sobkowiak}}]{Wil.ea:22}
{Wilawer}, E., {Oszkiewicz}, D., {Kryszczy{\'n}ska}, A., {et~al.} 2022, \mnras,
  513, 3242

\bibitem[{{Zappal{\`a}} {et~al.}(1990){Zappal{\`a}}, {Cellino}, {Farinella}, \&
  {Kne\v{z}evi{\'c}}}]{Zappala1990}
{Zappal{\`a}}, V., {Cellino}, A., {Farinella}, P., \& {Kne\v{z}evi{\'c}}, Z.
  1990, \aj, 100, 2030

\end{thebibliography}
\end{document}